 \documentclass{article}
   
 \usepackage{authblk}

\usepackage{lineno,hyperref}
 
\usepackage{graphicx,color}
    
\usepackage{amsmath}
\usepackage{amssymb}

\def\EE{\mathbb{E}}

 \newcommand{\rc}{\textcolor{black}}

\newcommand{\ea}{\end{eqnarray}}  
\newcommand{\ba}{\begin{eqnarray}}  
\newcommand{\ee}{\end{equation}}  
\newcommand{\be}{\begin{equation}}  
\newcommand{\een}{\end{equation*}}  
\newcommand{\ben}{\begin{equation*}}  
\newcommand{\ean}{\end{eqnarray*}}  
\newcommand{\ban}{\begin{eqnarray*}}

\DeclareMathAlphabet{\itbf}{OML}{cmm}{b}{it}

\newcommand{\dessous}[2]{
\renewcommand{\arraystretch}{0.5} 
\begin{array}[t]{c}
{#1} \\
\scriptstyle
{#2}
\displaystyle
\end{array}
\renewcommand{\arraystretch}{1.0}
}

\title{ Chaos and Order in the Bitcoin Market }

\author[1]{Josselin Garnier}
\author[2]{Knut Solna}
 
\affil[1]{\footnotesize Centre de Math\'ematiques Appliqu\'ees, Ecole Polytechnique, 91128 Palaiseau Cedex, France}
\affil[2]{\footnotesize    Department of Mathematics,  University of California, Irvine CA 92697}

\begin{document}
 
 \maketitle

\begin{abstract}   
   The bitcoin price has surged in recent years and it has also exhibited 
   phases of rapid decay. 
    In this paper we address the question to what extent this novel cryptocurrency  market
    can be viewed as a  classic or  semi-efficient market.  Novel and robust tools
    for estimation of multi-fractal properties are used to show that the bitcoin
    price exhibits a very interesting multi-scale correlation structure. 
    This structure can be described by a power-law behavior of the variances of the returns as   functions of time increments and it can be characterized by two parameters, the volatility and the Hurst exponent.
    These power-law parameters, however, vary in time.
    A new notion of generalized Hurst exponent is introduced which allows us to check
    if the multi-fractal character of the underlying signal is well captured. 
    It is moreover shown how the monitoring of the 
     power-law parameters can be used to identify regime shifts for the bitcoin price.  
    A novel technique for identifying the regimes switches based on
    a goodness of fit of the local  power-law parameters is presented. 
    It automatically detects dates  associated with
     some known events in the bitcoin market place.   
    A very surprising result is moreover that, 
    despite the wild ride of the bitcoin price in recent years and its multi-fractal and non-stationary
    character,  this  price has both local power-law behaviors and a very orderly correlation structure when it is observed on its entire period of existence.  
 \end{abstract}

   
 \section{Introduction}    
  
 Bitcoin is  the main cryptocurrency  and has been the subject of much interest,
 both from the point of view of speculation as well as
 from the point of view of the technology used in this market.
 In view of the success of the bitcoin market  
 other cryptocurrencies have tried to copy their technology, but 
 bitcoin remains the dominant technology with a market capitalization 
 of   about \$40  billions in mid-2017. 
    It is also interesting  to  understand  this new market from a price 
 or time series 
 modeling viewpoint. It is the latter question that motivates this paper.   
 
The bitcoin currency was launched  in 2009.  As a cryptocurrency  it is not 
regulated via centralized banks, but rather  transaction happens
over  decentralized computer networks and organized via 
block chain technology \cite{naka}.
In the early years  (2010-2013) bitcoin exchange was handled by
 Mt. Gox, an administration system based on Shibuya,
Japan, handling about 70\% of the transactions in late 2013. \rc{ Then in 
 early  2014 it became clear that   Mt. Gox had been  
 hacked resulting in  a loss  of bitcoins 
 valued at about \$450  millions.   
 Complaints about delays in withdrawing cash  from Mt. Gox 
 were mounting in early 2014  and on 
 7 of February 2014  Mt. Gox halted all bitcoin withdrawals
 citing a bug in the bitcoin software that made hacking possible.  }
 The subsequent loss of confidence in the 
 currency led to a rapid price drop. The confidence has, however, seemingly 
 rebounded as several  new exchange platforms have opened and the price surged 
 to  a high in early  2018 before a downward correction in the following months.
 
 One may expect that the lack of centralized bank  and regulatory agency
 means that the currency  is very sensitive and volatile and not efficient
 in a classic sense. Moreover, one may expect that the price evolution  after 
 the announcement of the hacking of Mt. Gox was special reflecting
 a reduced confidence in this currency  which may have persisted
 for some time.  These are among the questions we want to examine in this 
 paper.   
 
 The bitcoin market being the main cryptocurrency market has indeed
 been the subject of much research recently. This type of technology 
 may change  fundamentally the  arena for financial transactions.
 In view of its unique role as a pointer to what may be  to come
 in   financial markets it is thus  of  great interest to see what kind of price 
 structure and dynamics one finds  in the bitcoin cryptocurrency market.  
 
 In \cite{bit_cite} the authors analyze the bitcoin price with respect to 
 multi-scale temporal correlation structure and relate this to the concept of chaos. 
 They identify a low price period
between  July 2010 and February 2013 and a high price period from
 February 2013 till October 2017, by using a decomposition based  on price
 level.  A  main conclusion presented in \cite{bit_cite} is that chaos
 is present  in the case of prices, but  not for the returns. 
 It is also found that heavy distribution tails are the main factor driving the 
 chaos measure. 
 The returns refer to the relative price changes over a certain 
 time interval which typically corresponds to the sampling interval. 
  The analysis in \cite{bit_cite} is carried out partly via 
  multi-fractal detrended fluctuation analysis where data over various
  scales and intervals are detrended  and the associated residual moments
 of different orders are calculated \cite{mdfa}.   
   A similar technique is used for cross-correlation analysis in
 \cite{bit_corr}.    In \cite{bit_window,bit_window3}  
  various  types  of  detrended fluctuations analysis are used, 
  however,  with a moving window to track  changes.
 \rc{
 The approach presented here regarding power law decay in time for correlations of returns 
 is different from the approaches cited above,  in particular in that we focus on the second-order moments,
 moreover, we do not carry out any detrending of the data. }
 
 \rc{
 We remark that a power law behavior of the bitcoin price was analyzed 
 from a different perspective in \cite{stanley}, where  the authors analyze
 a power law, or relatively slow,  decay in the marginal distribution of the returns. 
 This interesting study suggests a universal behavior and a power 
 law decay of the marginal distribution with an exponent approximately equal to $2.5$.
 This value corresponds to heavier tail than in classic financial markets
 which typically exhibit an inverse cubic law of decay for the tail of the marginal distribution.     
 Importantly this means in particular that the returns have a finite second moment.
 Here we discuss   the issue of the marginal distribution of the returns in \ref{app:marg}
 where we show that our results are robust with respect to the presence
 of non-Gaussian returns distribution. 
 }
 
 In our analysis we use  an approach
 with a moving time window to track changes in the correlation structure.  
  We focus on the correlation structure
 of the returns and  set forth an approach
 based on  this structure  for detection of regime shifts.
 When we look at the returns in a time window we find 
 a very interesting and orderly scaling of the second-order moments of the returns
 as functions of return time increments.  
 The approach for segmentation is based on the residual relative to a fitted power law.  
 What is quite surprising in our analysis is that the local power
 laws coexist with a global power law also on the entire period of bitcoin existence.
  In the  classic Black-Scholes framework  \cite{booksv} 
 the log-price is a standard  Brownian motion  so that the returns are stationary,
 independent, and Gaussian.  We remark that  under the risk-neutral or pricing 
 measure,  the drift of the log-price is the risk free interest rate which
 follows from a no-arbitrage or efficient market condition.   
 Our focus here is, however, on the fluctuations of the returns,
 the Brownian part,  whose statistical structure is the same under the
 risk-neutral measure and the physical measure corresponding to 
 observed prices as considered  here. 
 To describe the considered power-law framework let $\Delta t$
 be the time increment  over which the returns are computed.  
We speak about power laws when the variances of the (zero-mean) returns
 are of the form $\sigma^2 |\Delta t|^{2 H}$, and we refer to $\sigma$ as the
 volatility and $H$ as the Hurst exponent.
  A classic model for such a power law is  
  fractional Brownian motion \cite{b18}. Indeed, this model gives 
 a generalization  of the classic Black-Scholes model  corresponding
 to the log return being standard Brownian motion with $H=1/2$. 
 In the fractional Brownian motion case the returns are not independent, consecutive  returns
 have a non-zero correlation coefficient  of $\rho_H=(2^{2H-1}-1)$. 
 The case $H<1/2$  corresponds to the anti-persistent case with
 negative correlation for the returns, while $H>1/2$ corresponds to the
 persistent case with positive correlation for the returns. As  
 observed early by Mandelbrot  \cite{mandelbrot71,mandelbrot97} it may
 however be appropriate  to model prices in terms of a local power-law process so that $H$ and $\sigma$ are time-dependent giving a 
 multi-fractional  Brownian motion in the Gaussian case.     Such type of mutifractal modeling has also been
considered for instance  in  equity markets \cite{bayraktar,cross},  in currencies 
markets \cite{oh}, in commodities market 
\cite{cajueiro04,alvarez02,elder08,jiang14}
and as a  model for physical measurements of various kinds 
\cite{temp,wind,PS,nuc}.     
     
     \rc{
 The analysis presented here of the bitcoin price identifies  four  main epochs. 
 As outlined above the epochs are identified by  minimizing  the residual between
 the empirical second-order moments and the modeled power laws, 
 with  the modeled power laws having constant parameters within each epoch. 
The first and the third  epochs are similar 
 in terms of persistent dynamics reflecting a strong-herding 
 behavior  or confidence in the market trend.  This corresponds
 to a super-diffusive behavior where the log-price changes grow
 superlinearly in terms of their second moments, so that the price
 can exhibit relatively larger price swings. 
  Wedged in  between these epochs is a period of much less persistence
 with a relatively low Hurst exponent close to a half correponding  essentially
 to an efficient market behavior.  
 Thus, this does not correspond to a 
 herding behavior with a  confidence in the market trend. 
 This mid-epoch starts approximately at the time that the bitcoin exchange
 Mt.  Gox was hacked.    We  find that in the third  epoch,
 after the confidence in the currency was reestablished, the persistence
 actually was slightly higher than in the  the first epoch of the bitcoin price
 path. This is an important observation that may explain the price surge.
 The fourth and final epoch corresponds to a Hurst coefficient of about .4
 and thus to an antiherding behavior in an epoch of relative strong price drops. 
It follows that for the bitcoin price  we may think 
of the Hurst exponent as a  market ``herding-index'',
a measure of a confidence in the market behavior.  Epochs of strong 
herding behavior are characterized by large Hurst exponents (i.e. larger than $1/2$).
We remark that  the epochs of strong persistence can be associated  with
relatively strong growth spurs in the price, 
  see Figure \ref{fig:price} below. }    
  
The estimation of the Hurst exponent tells us at  which level of market confidence
we are. It makes it possible to understand if a change in price level
and trend signifies a new market regime or if it can be seen 
as a random local price correction. 
From the financial viewpoint it is important to observe that the Hurst exponent
seems to be a better indicator of market confidence than the volatility
which classically is the most important financial parameter.     
Our  story does not stop here with a  time-varying market 
persistence.    What is very striking is that the market behaves
as if there were  an  ``invisible hand'' controlling the 
variations of the  persistence to generate
a   beautiful effective power law over the entire period of bitcoin existence,
see Figure \ref{fig:global} below.   Indeed the market is driven and controlled 
by the market participants so that on any given time epoch there
is an effective mean return  valid also on the subscales within this 
period.  For the bitcoin market over the total period considered this
mean return corresponds to a Hurst exponent of $H=.6$ corresponding
to a persistent market and positive correlation of the returns.

Our point of view in this paper is that the main quantity of interest is  
 the second-order correlation structure and what it tells
 us about market persistence and volatility at different scales and
 how these concepts are connected. 
 We remark that analysis of  high order-moments
 is   theoretically interesting, but becomes very sensitive to tail behavior 
 of returns   which limits their practical applicability. By focusing
 on the second-order temporal correlation structure  
 we can more robustly identify the quantities that are of direct
 financial interest. Moreover, we show that our multiscale analysis is robust 
 with respect to the marginal distribution of the returns.

Some technical points are discussed in the appendices:  In \ref{app:mBm} we give
the rigorous model for the multi-fractal process and relate it to our modeling of the
observations. The main tool we use to estimate the multi-fractal character 
of the bitcoin price is the scale spectrum and the associated  technique for  estimation of the
 power-law parameters.  We present the details of this concept and the estimation procedure
in \ref{app:S} in a way that can be easily reproduced.  
As mentioned we use a moving window to track the multi-fractal variations in
the price process. One may then wonder whether there is significant 
residual multi-fractality within the window.  In \ref{app:mom} we introduce
a novel notion of a generalized Hurst exponent and
a new method to check for within window multi-fractality. Using this  method we
find that in the epoch just after the Mt.  Gox  hacking there is significant
residual multi-fractality, but otherwise not.  One may further wonder whether 
the fact that the marginal value distribution of the returns associated with
the bitcoin price deviates from the Gaussian distribution is important,
in particular, whether  it could be a source for the observed multi-fractal character.
We show in \ref{app:marg} that this is not the case  by using a technique based on a Gaussian transformation.
Finally, we comment on a relation to classic chaotic systems in \ref{app:chaos}.
We show that indeed the breathtaking growth that the bitcoin cryptocurrency  has experienced
 cannot be well modeled by classic chaotic systems, in fact one has
 to model in terms of long-range processes of the type  considered
 in this paper.         
   
 \rc{The outline of the main part of the paper is as follows: In Section \ref{sec:model} we discuss
 the modeling and estimation procedure that we use for the power law scale spectrum
 and the results for the bitcoin data. 
 In Section \ref{sec:seg}   we present the approach for segmentation. 
 We consider measures  of scale-based correlations in Section \ref{sec:corr} and
 a measure of multifractality in Section \ref{sec:multi}. 
Finally,  in Section \ref{sec:concl} we conclude.}

 \section{Scale Spectrum of Bitcoin}
 \label{sec:model}
  
 \subsection{Scale Spectrum and  Power-law Parameters Estimation}
 \label{sec:data}
  
 We describe in this section how we compute the scale spectrum
 and  the estimates of the local power-law parameters:
 the Hurst exponent and the volatility.  The details are given in 
 \ref{app:S}.   
The data are the daily bitcoin prices in \$ denoted by
\ban
     P(t_n), \quad n=1,\ldots, N ,
 \ean
 where $t_n=t_1+(n-1) \Delta t$ and $\Delta t$ is the sampling rate (one day). 
We will consider a window of length $M$,  which below will be chosen 
as one year.
We denote the log prices
in the $k$th window (with $k\in \{1,\ldots,N-M+1\}$)   by
\ba
\label{eq:data}
    \big(a_0^{(k)}(i)\big)_{i=0}^{M-1} = \big(\log(P (t_{k+i})) \big)_{i=0}^{M-1}  .
\ea   
Here the window center time is $\tau_k=t_k+ (M-1)\Delta t/2$.
 The motivation for this notation is that we view these data   as
 the Haar  coefficients at level zero.    
 We next compute the continuous transform Haar
wavelet  detail  coefficients  at the different levels $j$ as in 
 (\ref{def:dji})   and the scale spectral data $S_j^{(k)}$ as in (\ref{eq:scs}). 
We use all available scales apart from the first one 
so that in  (\ref{eq:inertial}) we have
$ j_{\rm i}=2$  and $ j_{\rm e}  = \lfloor M/2 \rfloor$.
Thus, we do not use the first scale which is most sensitive to ``measurement noise''. 
Indeed, it is seen from Figure \ref{fig:global} below that the first scale spectral
point is slightly enhanced relative to the fitted model. 
   
  If the observations come from
a fractional Brownian motion with constant Hurst exponent $H$   and volatility
$\sigma$ 
in the way described in \ref{app:mBm}
then we have 
\begin{equation}
\EE\big[ S_j^{(k)} \big] = \sigma^2 h(H)  2^{j(2H+1)} ,
\label{def:modelpl}
\end{equation}
for 
\begin{equation}
\label{hfunc}
h(H) =  \frac{(1-2^{-2H})}{(2H+2)(2H+1)} ,
\end{equation}
so that 
$$
\log_2\EE \big[{S}_j^{(k)} \big]  =   \log_2(\sigma^2 h(H)) +  j(2H+1) .
$$
We can then obtain the estimates  of the Hurst exponent 
$\hat{H}(\tau_k)$ 
and the volatility $\hat{\sigma}(\tau_k)$ by a
 least squares procedure   
(linear regression) of $ \log_2(S_j^{(k)})$ as shown in \ref{app:S}.  
   We remark that the estimated volatility is the volatility on the $\Delta t$ time scale.
The local volatility on the time scale $\tau=m\Delta t$ is 
$ \widehat{\sigma}(\tau_k)   m^{\widehat{H}(\tau_k)} $.

 \subsection{  Bitcoin Multi-scale Structure }
 
\begin{figure}[htbp]
 \begin{center}
\begin{tabular}{c}
\includegraphics[width=6.0cm]{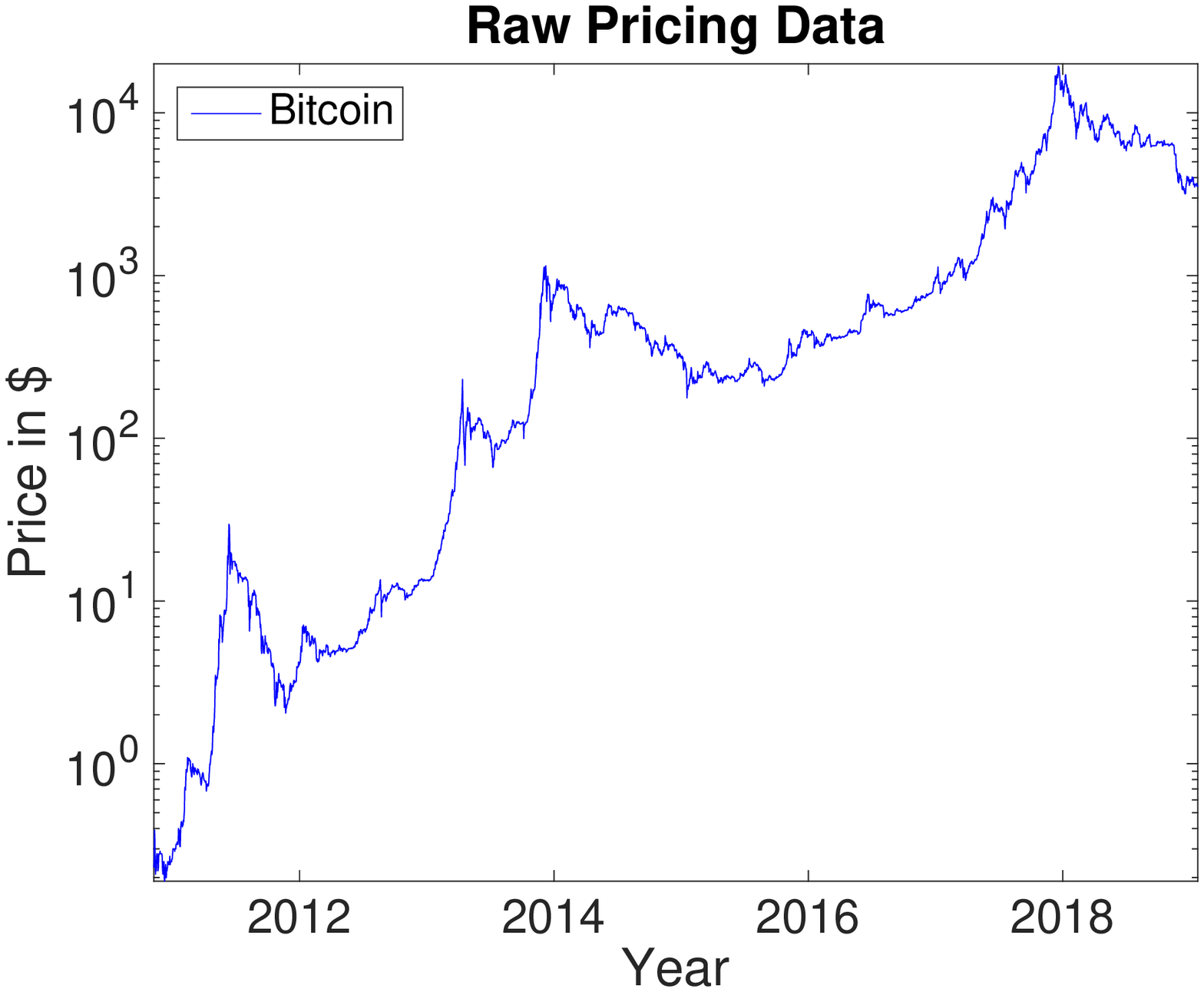}
\includegraphics[width=6.0cm]{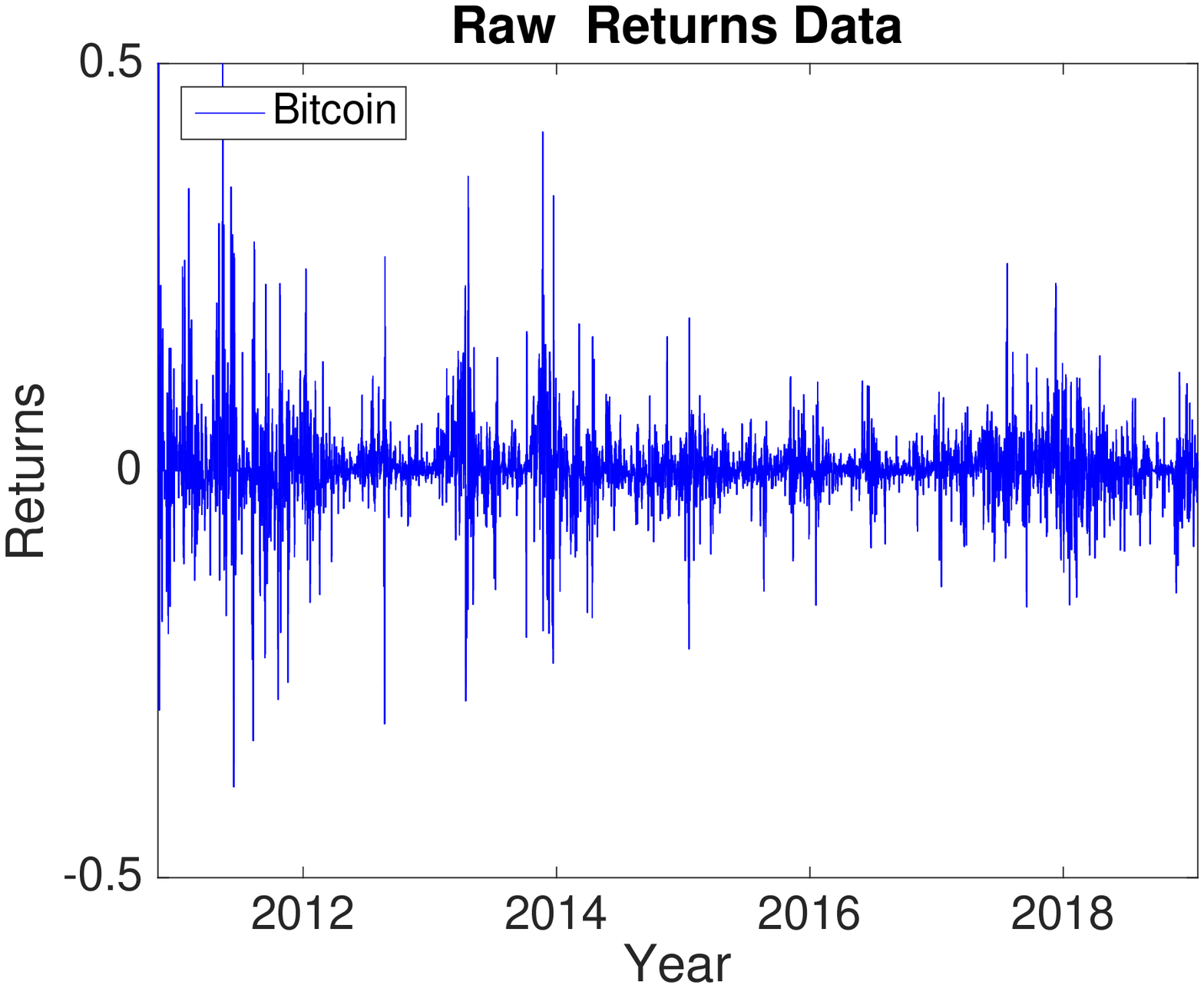}
 \end{tabular}
    \caption{  The daily raw bitcoin price in log scale (left) and the associated returns, or relative price
    changes (right). 
   }
   \label{fig:price}
\end{center}
\end{figure}

 \rc{ In Figure \ref{fig:price}  the left plot shows the daily bitcoin price (in log scale) in~\$, 
 that is, $P(t_n), \quad n=1,\ldots,N$,
  with $t_1$ being November  5th 2010,   
  $t_N$ being  January  23rd, 2019, and $N=3000$.
There is a price observation for every day of the year.
  The data were obtained from the web site 
  {\it www.coindesk.com/price/bitcoin} as daily closing prices.  }
 We see that after its introduction the bitcoin experienced a very rapid growth,
 which was somewhat tamed after about two years. However, the growth
 stagnated around the beginning of 2014 with the hacking of Mt. Gox exchange
 until about the beginning of 2017 when a period of strong  growth culminated 
 in  a maximum  price in early 2018. 
 The right plot in Figure \ref{fig:price}  shows the returns, that are
 \ban
   R_n= \frac{P(t_{n}) - P(t_{n-1})}{P(t_{n-1})}  .
 \ean
 From 
 the plot it appears that the volatility was highest in an initial phase 
 and that it has rebounded somewhat in recent years. We examine
 more closely these qualitative observations by looking at the estimated local
power-law parameters next. 
 
\begin{figure}[htbp]
   \begin{center}
\begin{tabular}{c}   
\includegraphics[width=6.0cm]{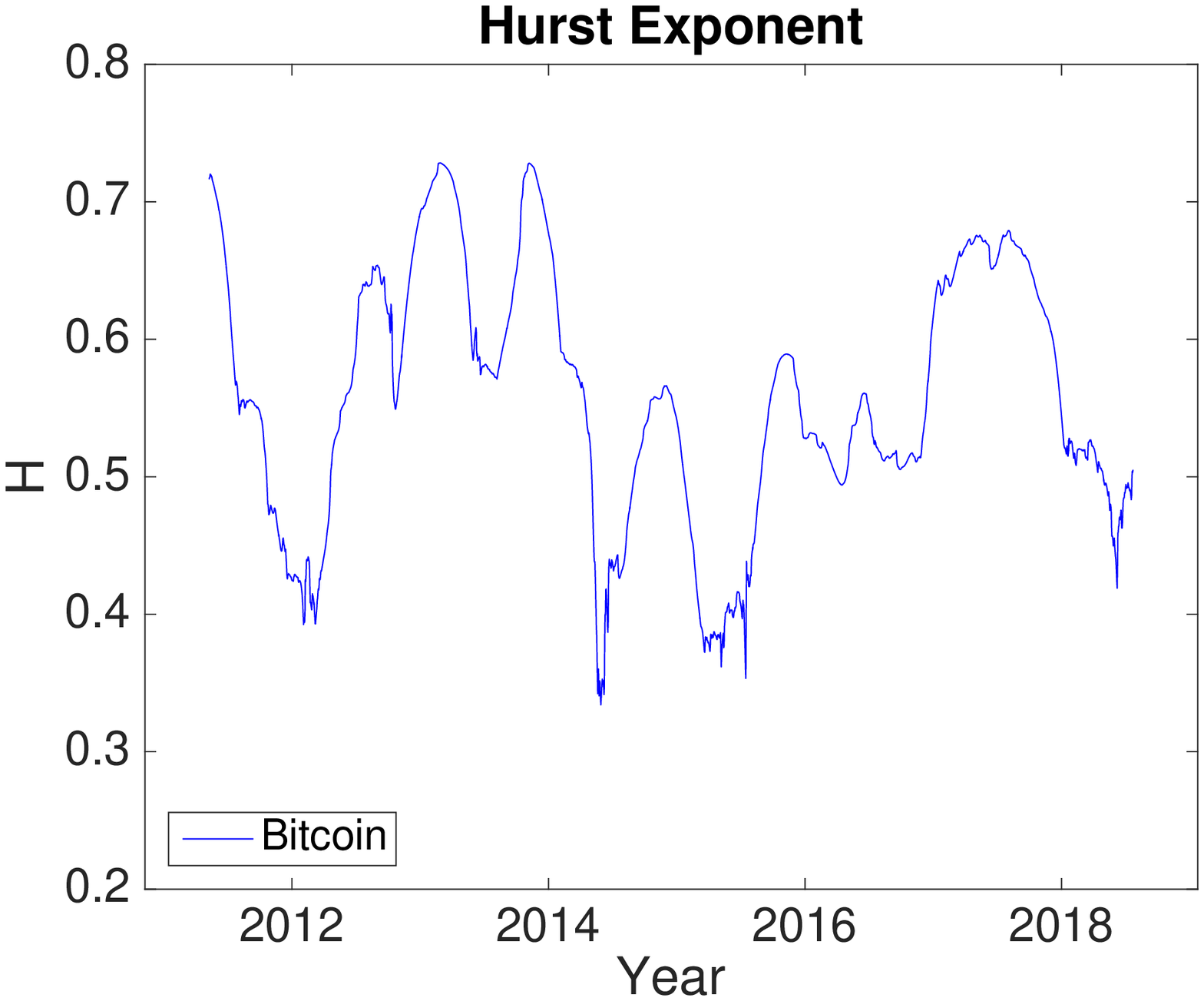}  
 \includegraphics[width=6.0cm]{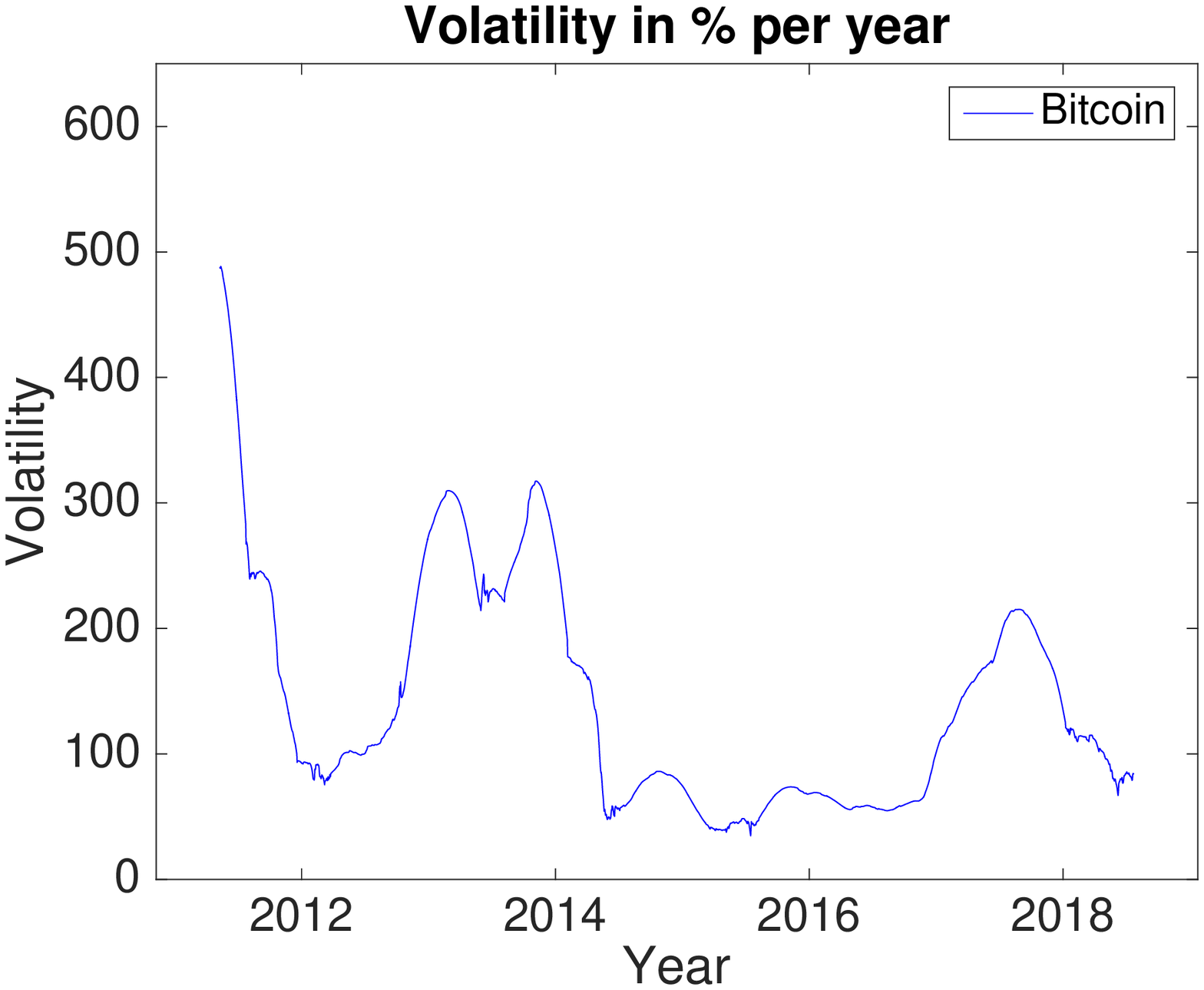}
 \end{tabular}
     \caption{   \rc{The  left  plot shows the Hurst exponent as function of the 
     window center time used in the estimation (thus the first data point
     is half a year after the time of the first  price observation with a   
     window width of one year year)}.
    The right  plot shows  the corresponding estimated  annual-scale volatility. 
        }
   \label{fig:pars}
\end{center}
\end{figure}

 In Figure \ref{fig:pars} we show the estimated local Hurst exponent, $\hat{H}(\tau_k)$,
 in the  left plot and  the  estimated local volatility, $\hat{\sigma}(\tau_k)$, in the 
right  plot. These estimates are obtained as described in the previous 
 section with a window width of one year used for computing the local scale spectrum. 
 We see that there are considerable variations in the Hurst  exponent.  
 In an early phase of the bitcoin existence both the volatility and the Hurst 
 exponent are high while they decrease after the hacking of Mt. Gox
 in early 2014. In the last years the Hurst  exponent has again increased
 and it is also seen that the price movements are larger in this period.     
 Note that the time period of minimum price after the hacking event occurs around 
 mid 2015, which is an epoch of relatively small price drift.  This epoch corresponds
 approximately with  the minimum Hurst exponent estimate.  
 There is   a brief epoch just after the Mt. Gox hacking with the 
 lowest  Hurst exponent and thus strong anti-herding behavior,
 a special event is indeed detected,   however its duration is 
  shorter than the window width of one year.
   
\begin{figure}[htbp]
   \begin{center}
\begin{tabular}{c}   
  \includegraphics[width=6.0cm]{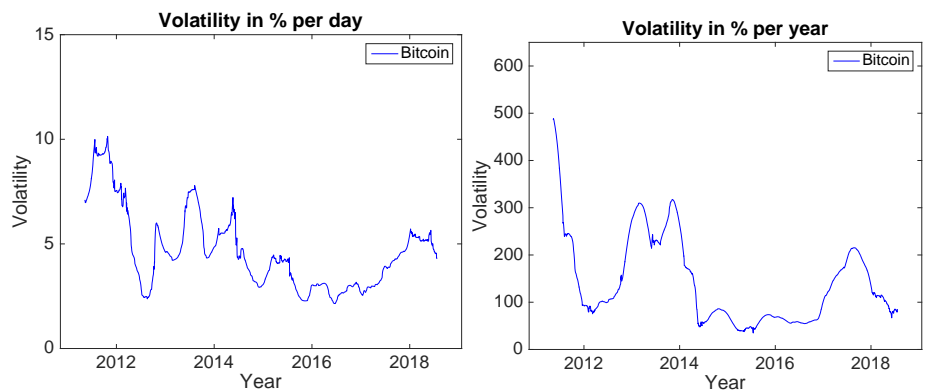}  
\includegraphics[width=6.0cm]{fig_masterS.eps}  
 \end{tabular}
     \caption{   The left plot shows the estimated volatility  on the daily scale,
          the right  plot shows the estimated volatility  on the annual scale. 
        }
   \label{fig:pars2}
\end{center}
\end{figure}

In Figure \ref{fig:pars2} we show in the right plot the volatility on the annual scale 
(as in the right plot of Figure \ref{fig:pars}) and
the volatility on the daily scale in the left plot. This illustrates the fact 
that the variations in the Hurst exponent are
the  primary  driver of annual scale volatility. 
From the financial perspective it is then clear  that variations
in the Hurst exponent are of primary importance as they largely drive
the annual scale volatility  which is the classic measure of risk.
In particular it indicates that a volatility estimate based on the
mean square of the returns, the empirical quadratic variation, may be strongly
biased.  

\begin{figure}[htbp]
  \begin{center}
\begin{tabular}{c}   
\includegraphics[width=6.0cm]{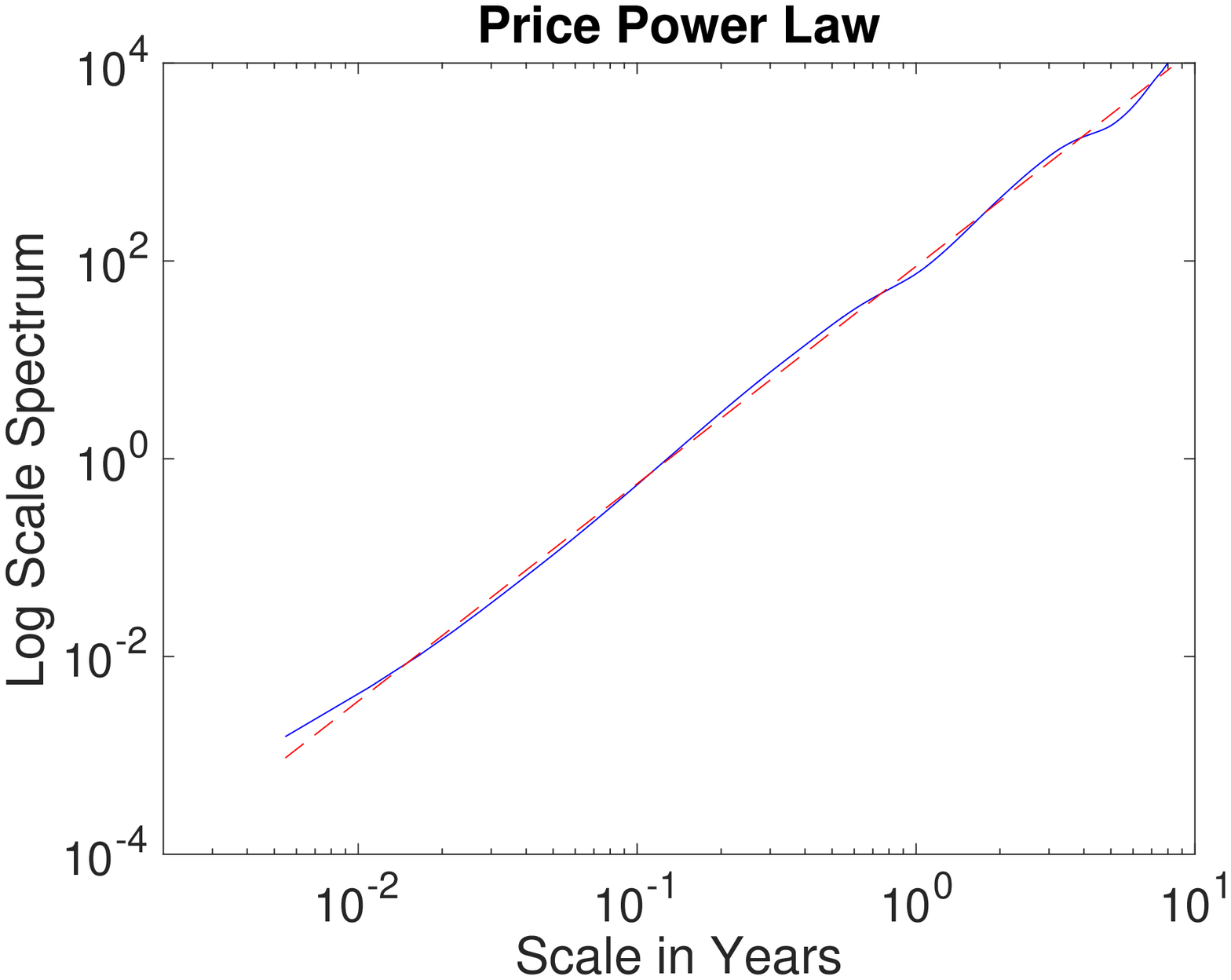}
 \end{tabular}
    \caption{   The global scale spectrum (blue solid line) estimated over the full data set. The estimated parameters are    $H=.60$, $\sigma=173\%$. The red dashed straight line is the power law spectrum with the estimated parameters.
   }
   \label{fig:global}
\end{center}
\end{figure}

Figure \ref{fig:global} shows the global power law, that is the case
when the we use all the data shown in Figure \ref{fig:price}  to compute
the scale spectrum.  We can observe a nice power law 
with effective parameters $H=.6$ and $\sigma=  178\%$ on the annual scale.  

We have seen that the spectral characteristics of bitcoin price show temporal 
variations, moreover, that they aggregate to form a nice power law when observed over the entire period of bitcoin existence.
It may then be natural to ask how one best can do a partial aggregation 
to naturally segment  the bitcoin price into sections where
the power law is approximately homogeneous within each section and
we discuss this in the next section.

\section{Segmentation and Regime Switch Detection}\label{sec:seg}

  Consider  a  partition of the full data set
 into $Q$ disjoint segments with width $M_q \Delta t$, $q=1,\ldots,Q$ (we have $\sum_{q=1}^Q M_q=N$). 
We apply the estimation procedure  described in \ref{app:S} on each segment.
We denote the estimated scale spectrum in each window
 by $S^{(q)}_j$, $j\in \{ j_{\rm i}^{(q)},\ldots, j_{\rm e}^{(q)}\}$, where
  $\{ j_{\rm i}^{(q)},\ldots, j_{\rm e}^{(q)}\}$ is the inertial range of interest (we take $j_{\rm i}^{(q)}=2$ and $j_{\rm e}=\lfloor M_q/2 \rfloor$).  We also denote the modeled power-law scale spectrum
with the estimated parameters (volatility $\sigma^{(q)}$ and Hurst exponent $H^{(q)}$)
by $\overline{S}^{(q)}_j$, $j=j_{\rm i}^{(q)},\ldots, j_{\rm e}^{(q)}$ (see (\ref{def:modelpl})).
  Then we define  the total spectral  residual   by
 \ba
 R (M_1,\ldots, M_Q) =     \sum_{q=1}^{Q}   \sum_{j=j_{\rm i}^{(q)}}^{j_{\rm e}^{(q)}}
     \frac{1}{j}   \left( \log_2\big(S^{(q)}_j \big)   -   \log_2\big(\overline{S}^{(q)}_j\big)  \right)^2  .
 \ea
 Note that 1) the weighting of the spectral residuals
is uniform with respect to the windows, which serves to penalize relatively short windows;
2) the weighting is proportional to the reciprocal  scale, $j^{-1}$,  which reflects
 the larger variance of the spectral data for longer scales. 
  Finally we estimate the optimal segmentation by minimizing
 $R$  with respect to the partitition $(M_1,\ldots,M_Q)$ via 
 an exhaustive search (the convexity of the function $R$ is not known).   
   
\begin{figure}[htbp]
  \begin{center}
\begin{tabular}{c}  
 \includegraphics[width=6.0cm]{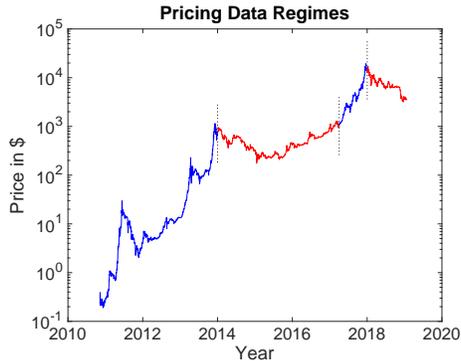}  
  \end{tabular}
\end{center}
    \caption{\rc{The segmentation of the price time series into four  epochs. 
    The vertical lines gives the change points in between segments. 
   }}
   \label{fig:seg}
 \end{figure}

\begin{figure}[htbp]
 \begin{center}
\begin{tabular}{cc}   
\includegraphics[width=5.0cm]{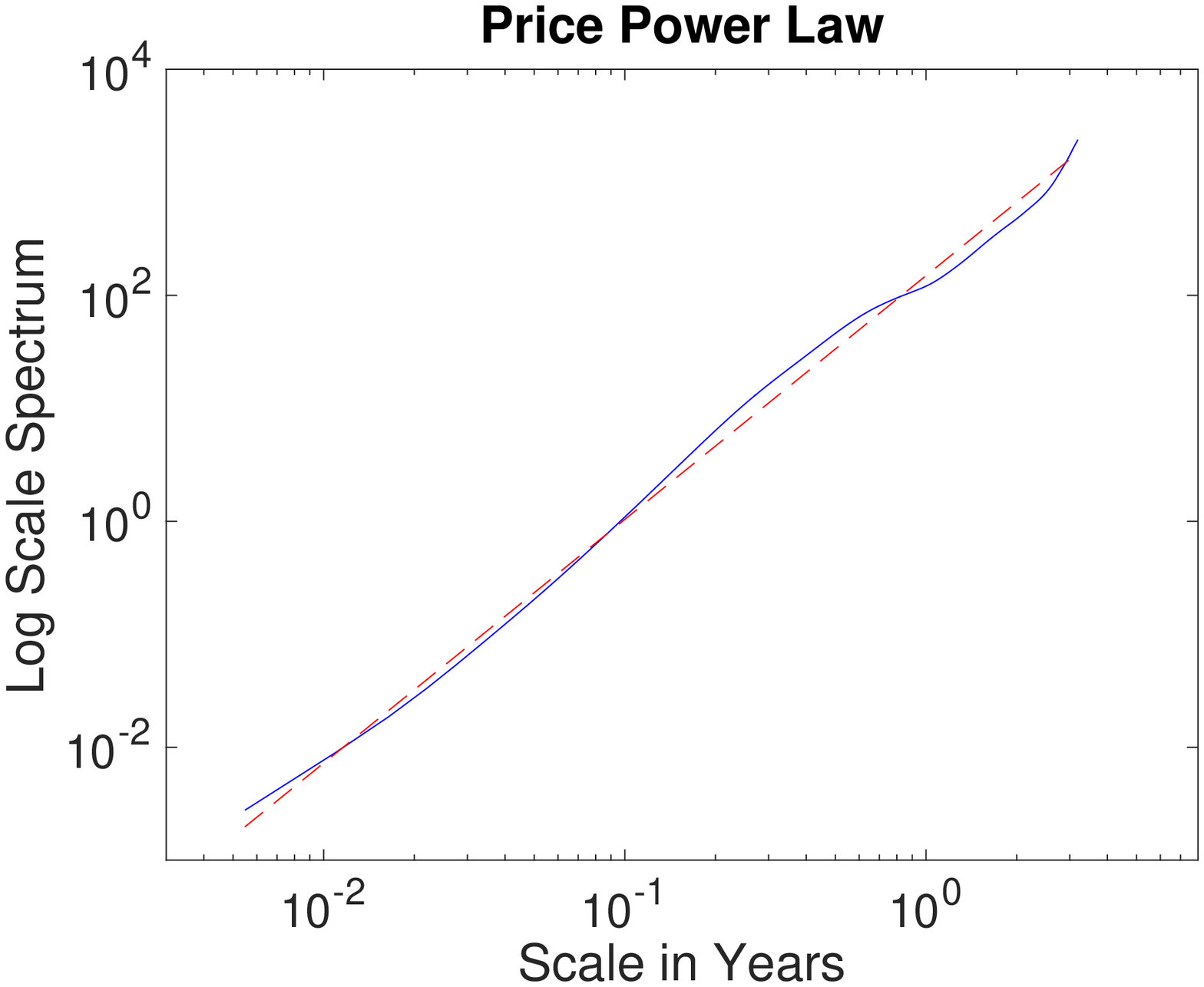}
\includegraphics[width=5.0cm]{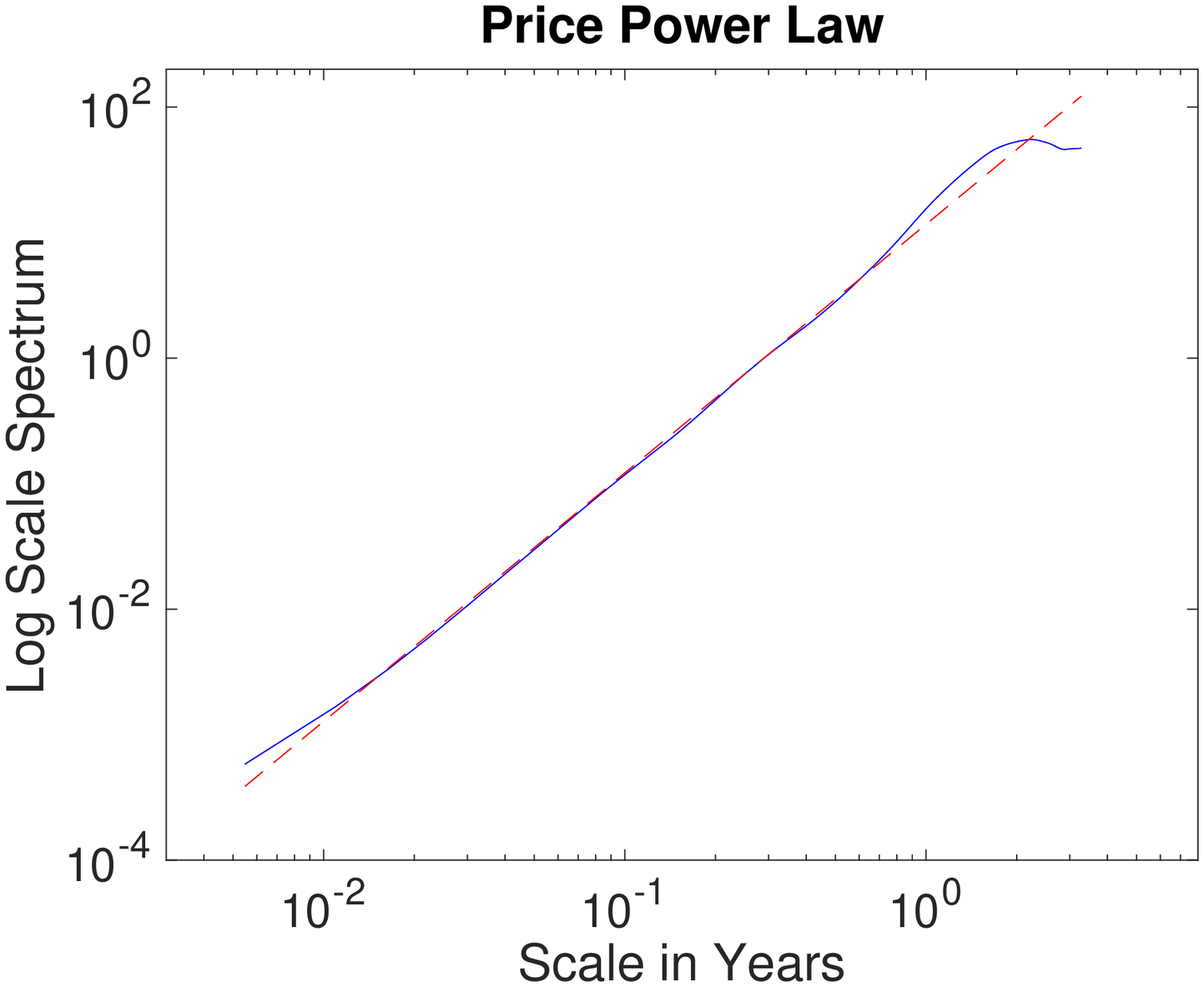} \\
\includegraphics[width=5.0cm]{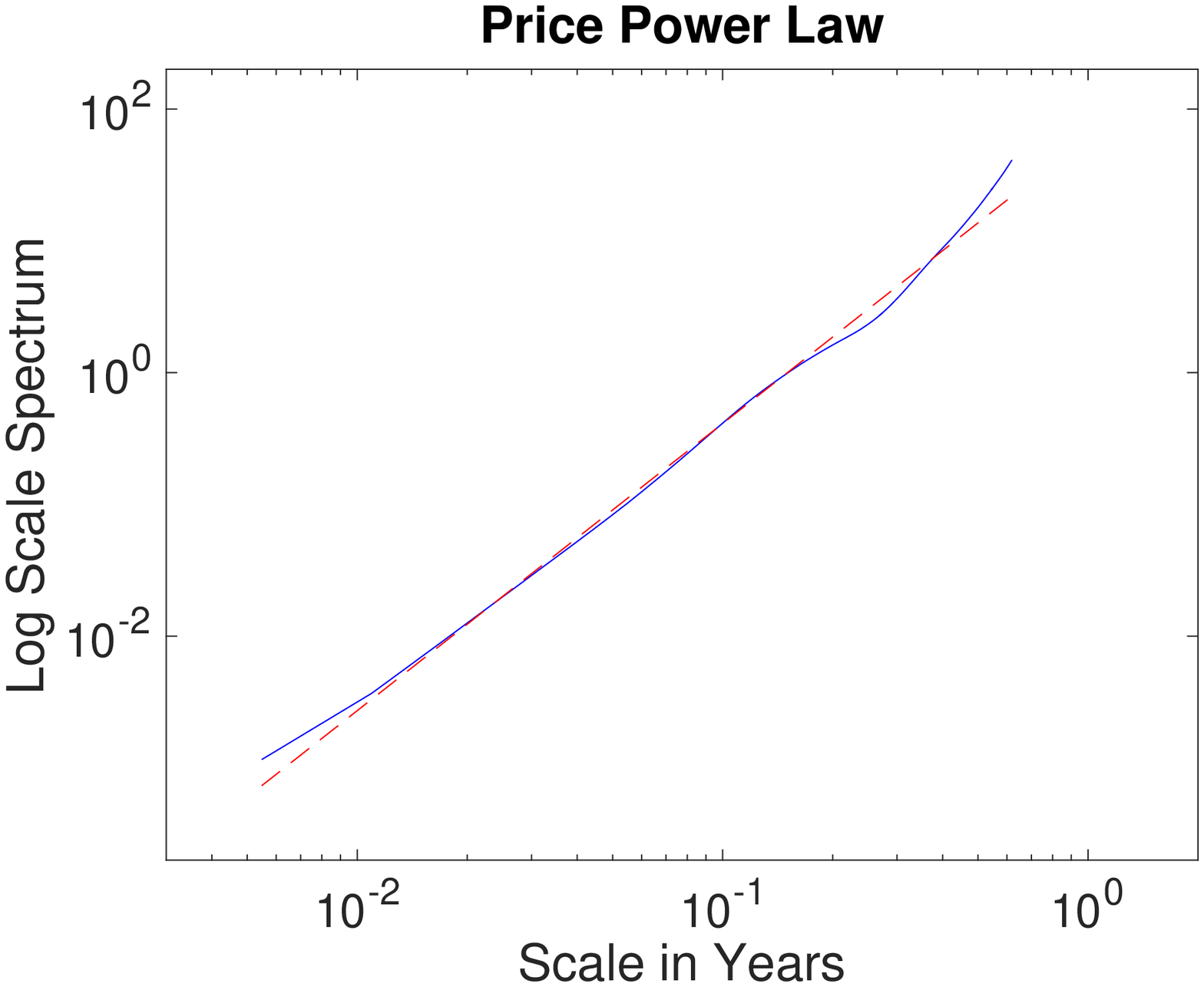}
\includegraphics[width=5.0cm]{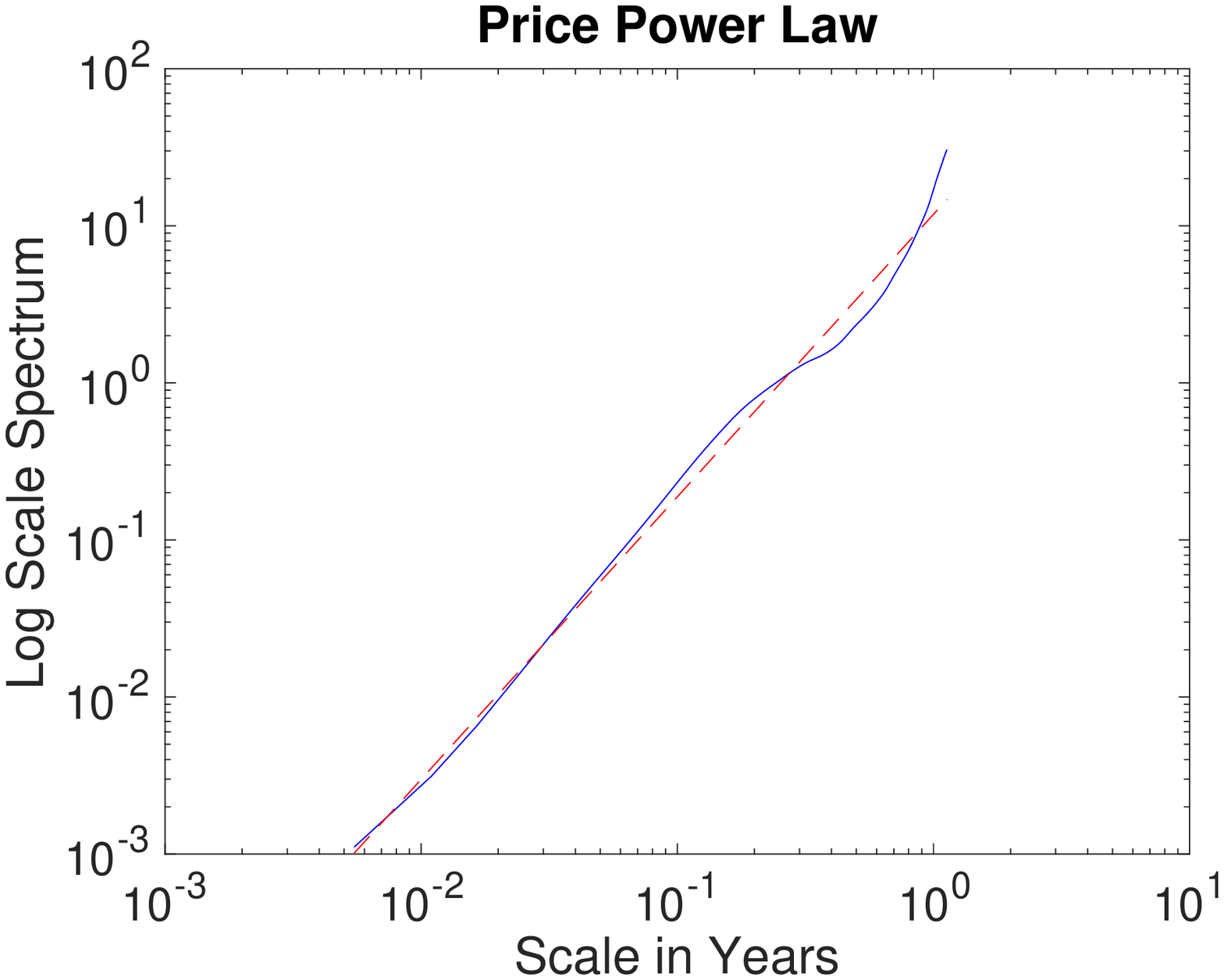}
 \end{tabular}
\end{center}
     \caption{  \rc{The scale spectra  in the four estimated epochs (blue solid lines) and the power law spectra with the estimated parameters (red dashed
     straight lines). 
     The estimated parameters are for the four epochs
    $H=.58, .49, .59, .40$ and  $\sigma=225\%, 62\%, 145\%, 62\%$. 
   }}
   \label{fig:seg2}
 \end{figure}

  \rc{In Figure  \ref{fig:seg}  we  show the result  
  of the segmentation procedure  when we let $Q=4$ and
 we implement a two-level search with a coarse grid size of half a year
  and a fine grid size of five days.
   The estimated power-law parameters are for the four segments
    $H=.58, .49, .59, .40$ and  $\sigma=225\%, 62\%, 145\%, 62\%$. 
  The first change point (January 16th 2014) corresponds to the hacking 
  of Mt Gox,  while the second change point (April 23rd 2017) corresponds
  to  the start phase of the second strong growth period of the bitcoin price 
  approximately at the time it reaches its previous maximum.
  The final change point (December 5th 2017) 
  corresponds to the initiation of a phase of declining 
  bitcoin price recordings.    }
  In Figure  \ref{fig:seg2} we show the scale spectra corresponding to the
  data in the four segments. 
  \rc{We remark that using  more than  four segments
  does nor lead to a further reduction in the residual $R$.   }  
  It is remarkable that the local spectra have different power law behaviors
and at the same time that the global spectrum has  also a power law behavior (Figure~\ref{fig:global}).
\rc{ The precision of the Hurst estimates  can be assessed by the precision obtained with fractional Brownian motion. For the window sizes of the data corresponding 
to  the four windows in Figure \ref{fig:seg2}, 
this gives a relative standard deviation for the Hurst exponent  that is approximately
7\% for the two first segments and 10\% for the last  two shorter segments.
Moreover, for the two first segments there is a negligible bias while
for the last two segments the bias is approximately  $-.03$.    
The  estimator is approximately Gaussian distributed, and thus the 
first, third, and fourth estimated values for $H$ can be said to be significantly different from $H = 1/2$. }

\section{\rc{Cryptocurrency Multiscale Correlations}}\label{sec:corr} 
 
\begin{figure}[htbp]
 \begin{center}
\begin{tabular}{c}   
\includegraphics[width=6.0cm]{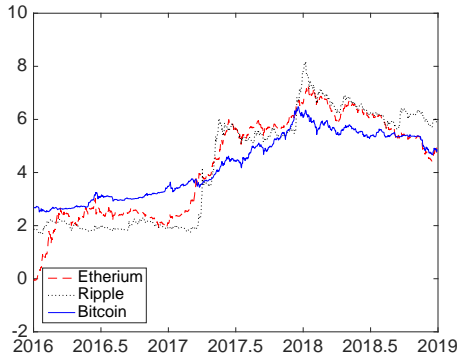}
 \end{tabular}
\end{center}
     \caption{\rc{Prices of the three cryptocurrencies considered; top (orange) line: bitcoin; middle (blue) line: Ethereum; bottom (red) line: Ripple. The vertical axis is in log scale, but with  shifts for the lines to a common mean level. 
   }}
   
   \label{fig:corrA}
 \end{figure}

\begin{figure}[htbp]
 \begin{center}
\begin{tabular}{ccc}  
\includegraphics[width=3.8cm]{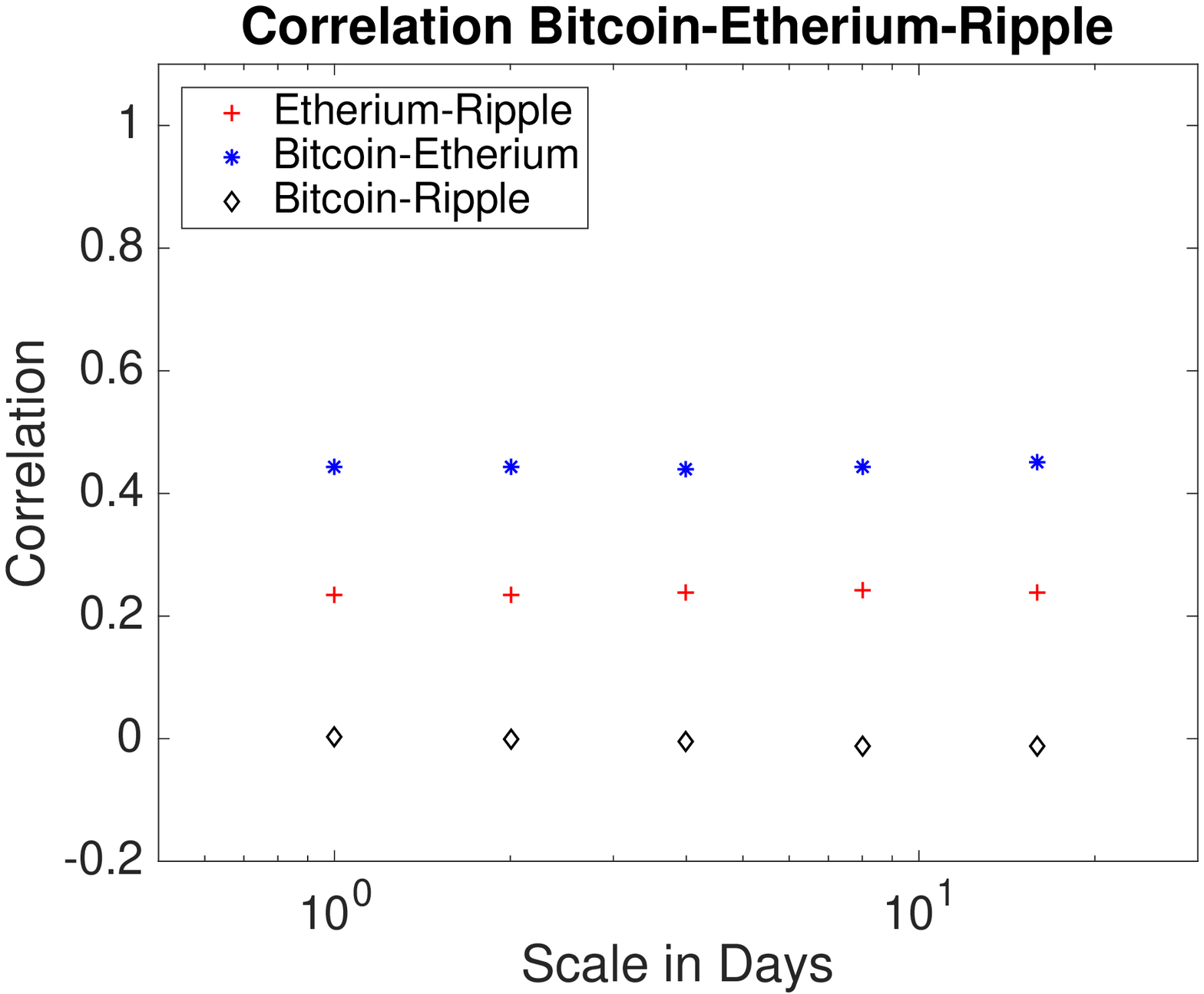}
\includegraphics[width=3.8cm]{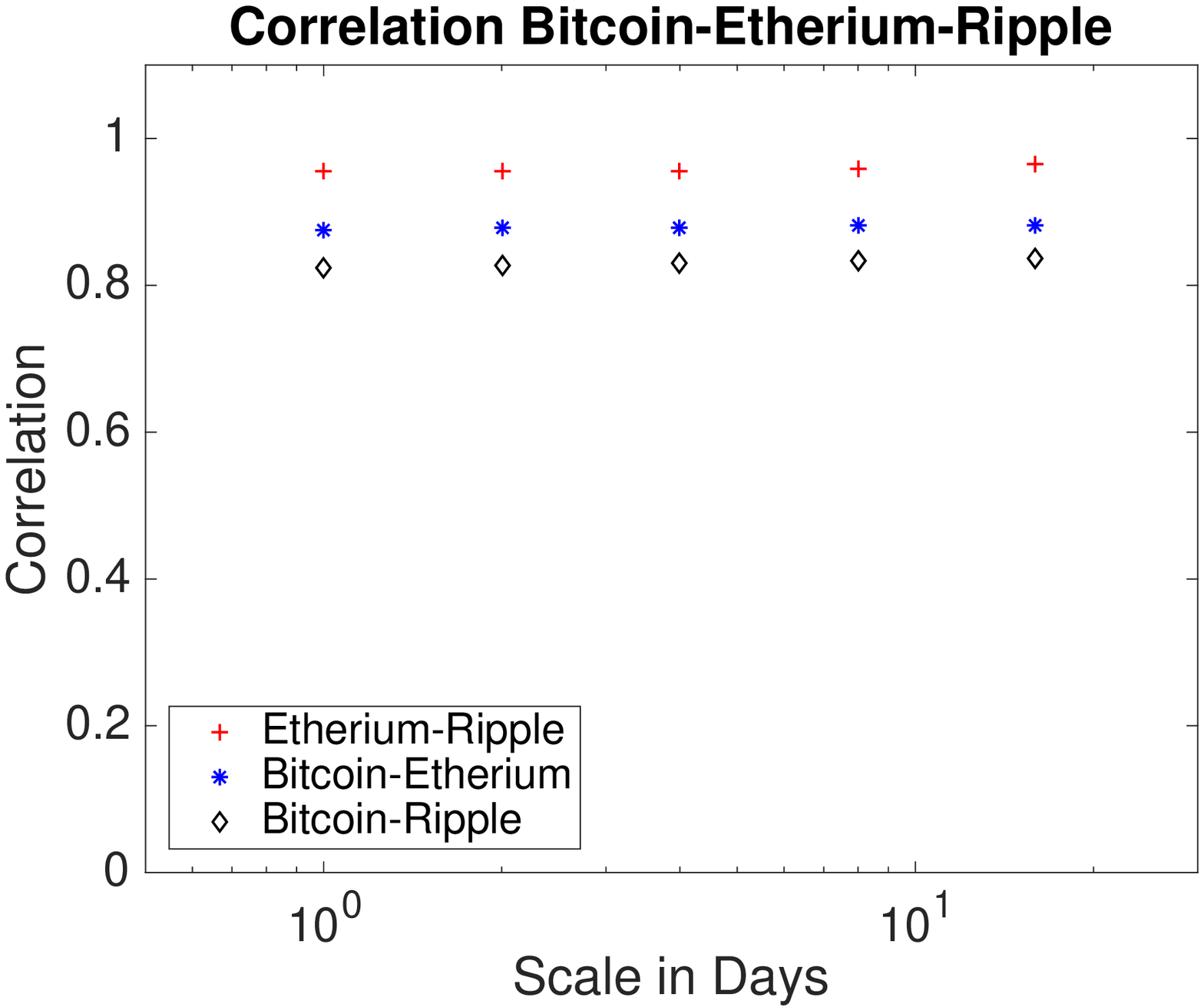}
\includegraphics[width=3.8cm]{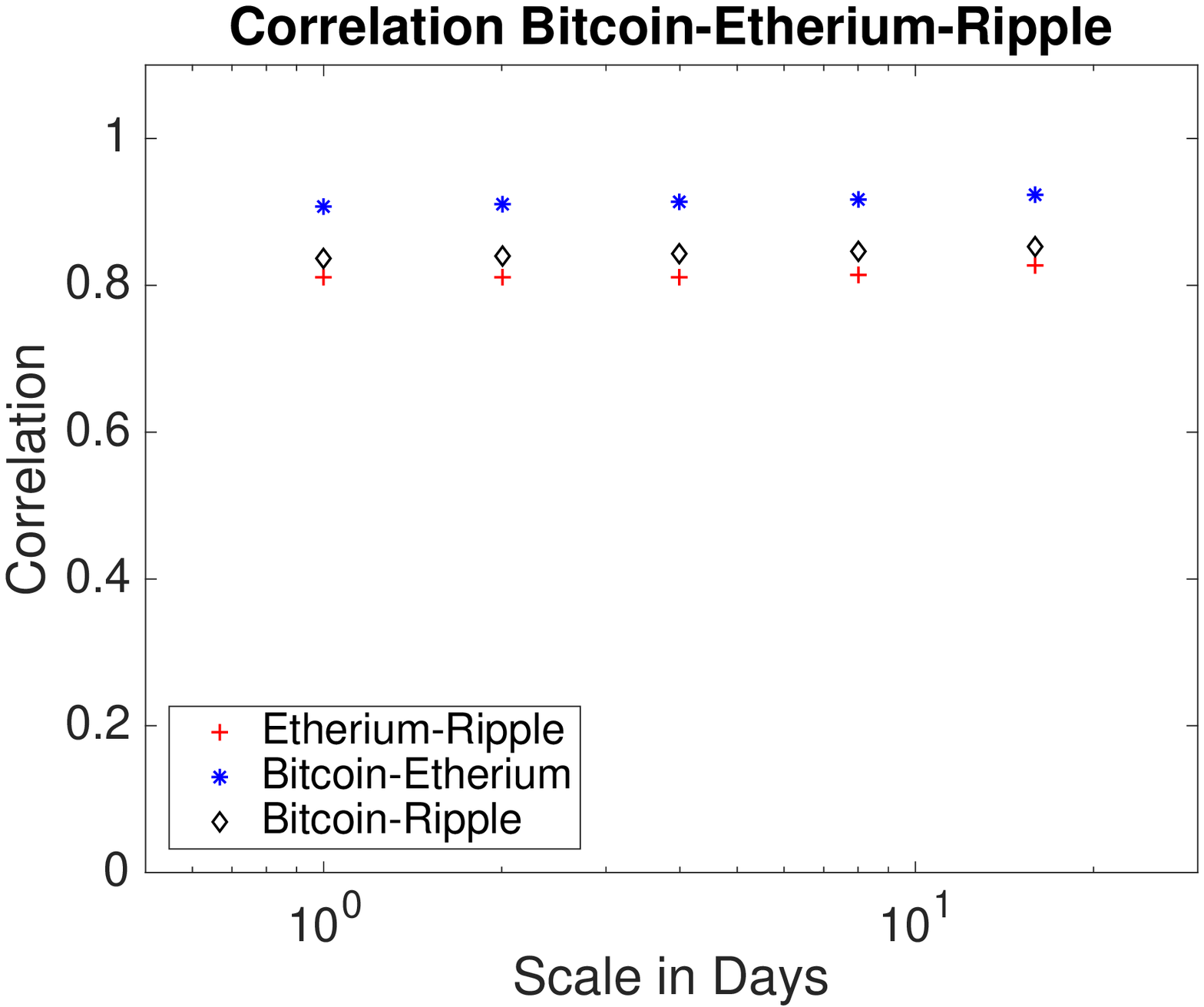} \\ 
\includegraphics[width=3.8cm]{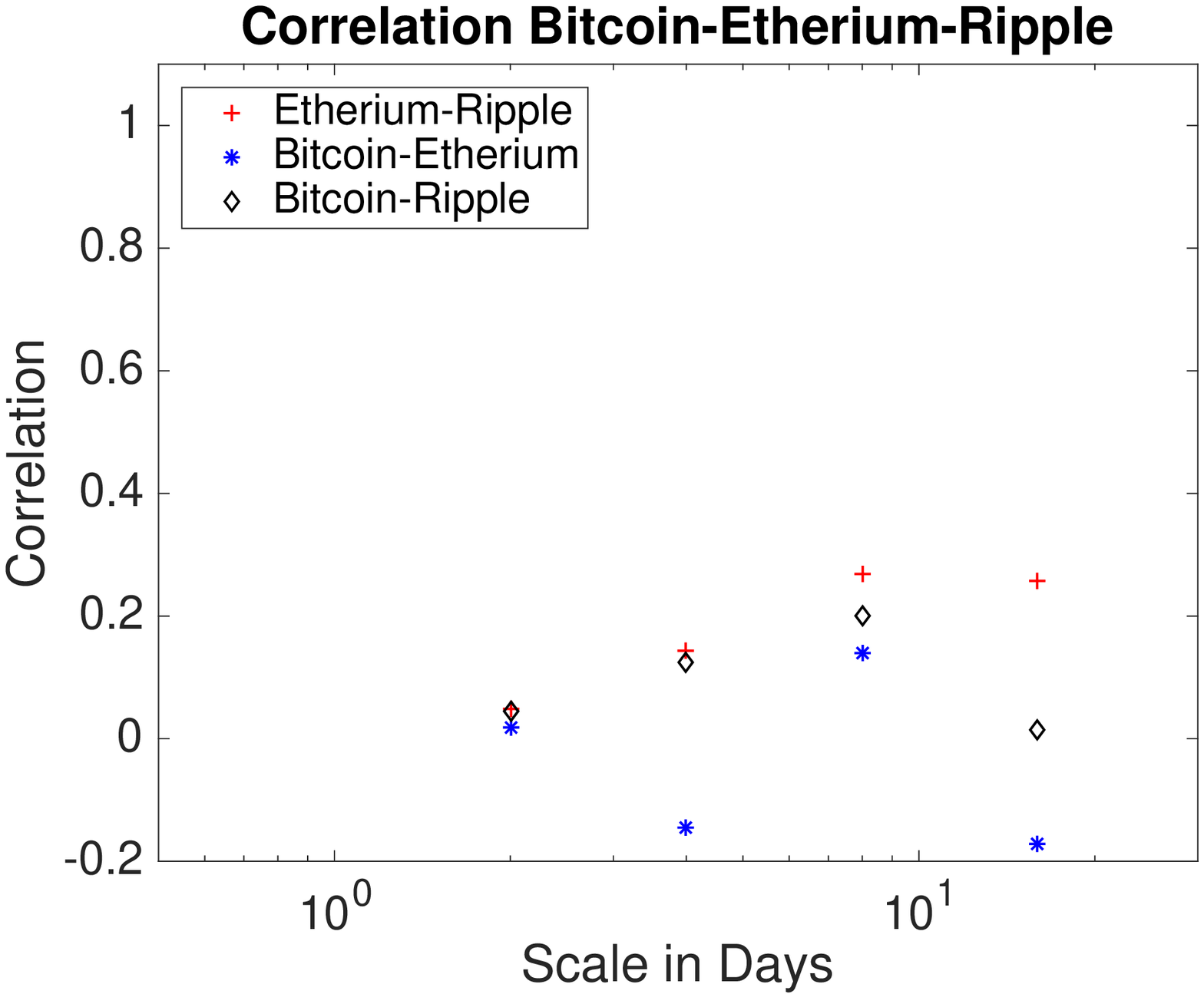}
\includegraphics[width=3.8cm]{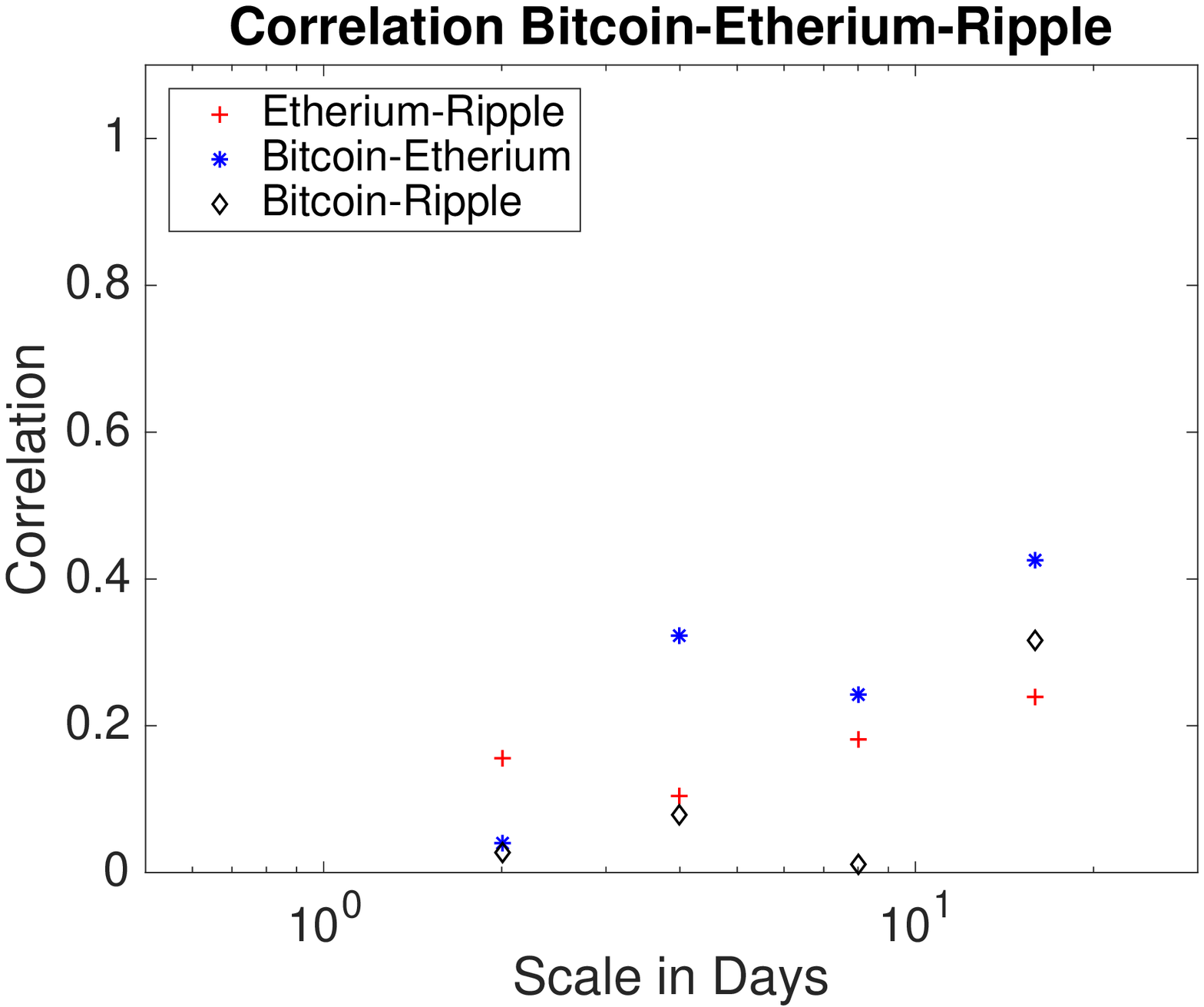}
\includegraphics[width=3.8cm]{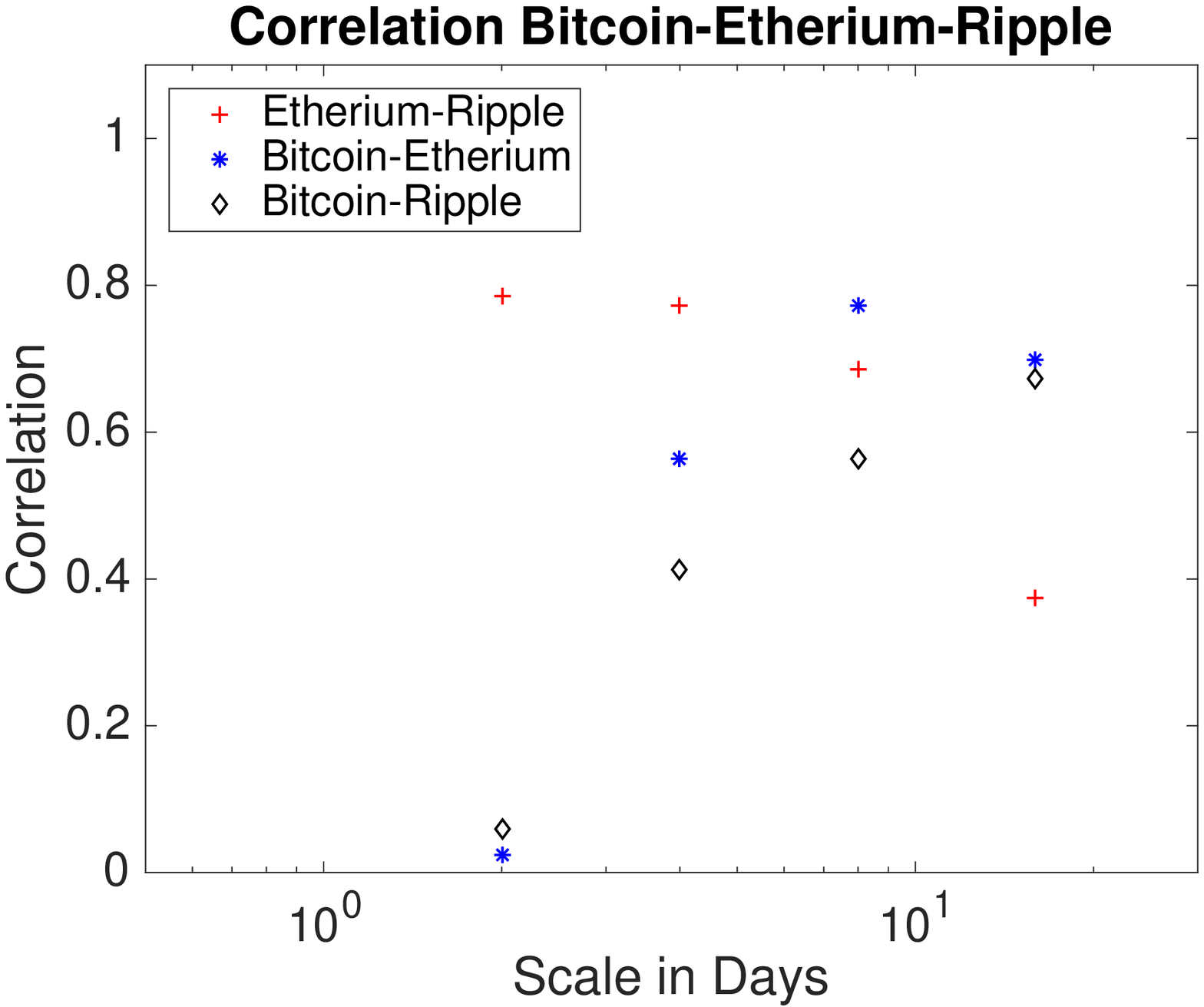} 
 \end{tabular}
\end{center}
     \caption{\rc{Top plots show the correlation coefficients of the approximation coefficients
     in the three years 2016, 2017 and 2018 (left to right), while the bottom plots
     show the correlation coefficients of the difference coefficients. 
The (red) crosses are the correlation coefficients 
between  Ethereum and Ripple,
the (blue) stars between Ethereum and bitcoin and the (black) diamonds between
Ripple and bitcoin.   
   }}
   \label{fig:corrB}
 \end{figure}

 \rc{We consider next how the bitcoin price is correlated with some
 other cryptocurrencies.  
 The two other cryptocurrencies considered are Ethereum  and Ripple.
 We show  in Figure \ref{fig:corrA} the three currencies
 (daily log-prices) on a common interval of existence,  
 the period January 1st 2016 till December 29th 2018. 
 We consider then the three periods corresponding to 2016, 2017 and 2018 
 respectively. In the context of the regimes for the bitcoin price
 estimated  in the previous section this corresponds respectively
 to epochs of (i) relative efficiency and small persistence
(ii) strong persistence and a growth spur (iii) strong antipersistence
and significant price drop. 
We compute first the correlation coefficients of
the approximation coefficients  of the log-price time series, 
that are the averages of the log-prices over 
intervals of length $2^j$  for  $j=0,\ldots,4$.
The approximation coefficients  give the price level variations at different scales, the dyadic scales 
between $1$ day and $16$ days. 
We remark that with one year of daily observations this corresponds
to a relative accuracy ranging from roughly 5\% till 20\%. 
The case $j=0$ corresponds to the classic measure of correlation 
of the time series. 
The results are shown in the top subplots of Figure \ref{fig:corrB}.
The first epoch of relative efficiency is associated with 
smaller cryptocurrency correlation. Moreover,  the bitcoin is consistently
stronger correlated with Ethereum than with Ripple and the 
correlations are stable with respect to scale. 
 The bottom subplots show the  correlation coefficients of the 
difference coefficients, that are the $d_j$ as in (\ref{ddef}) at levels $j=1,\ldots,4$.
The case $j=1$ corresponds roughly to daily returns, while
the other scales give measures of returns for longer scales. 
In general the correlation coefficients for the difference coefficients
are smaller than for the approximation coefficients and  
the variability with respect to scale is higher. 
Indeed the correlation coefficients  in the first epoch corresponding to
relative efficiency are  small. 
In general there is
an enhanced correlation with respect to scale for the correlations
between bitcoin and the two  other currencies in the strongly persistent or
anti-persistent epochs, while there
is little observable systematic variation with respect to scale for
correlations between Ethereum and Ripple.  }


\section{\rc{Bitcoin Multifractality}}
\label{sec:multi}
\begin{figure}[htbp]
   \centering
       \includegraphics[width=6.0cm]{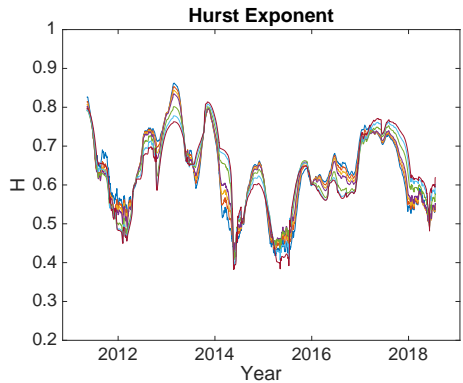}  
   \caption{Generalized  Hurst exponent  
   estimates for $q \in \{ 1/4, 1/2, 3/4, 1, 2, 3, 4\}$.
   }
   \label{fig:moment}
\end{figure}

\rc{
We briefly comment on measure of local multifractality for the bitcoin price
and discuss this in more detail in \ref{app:mom}.
We generalize the  scale spectrum in   (\ref{eq:scs})   to the $q$-moment scale spectrum,
$S_j(q)$ defined in (\ref{scale-sp:p}). 
We can  then,  as in the above case  with $q=2$, estimate the Hurst exponent
via linear regression with respect to scale of the log-scale scale spectrum.
In the case that the underlying process is fractional Brownian motion and we
have a long data set   this estimate is independent of $q$ while
strong $q$-dependence is a measure of multifractality. 
We show the result of the Hurst exponent estimate for various values of $q$
in Figure \ref{fig:moment}. We note that  the epoch associated with the Mt. Gox is associated
with a particular high degree of multifractality. Thus, this measure identifies 
the special epoch associated with the hack which not only distinguishes 
itself via a change in  Hurst exponent, but also a brief epoch of deviation from 
``pure'' fractional scaling  structure even locally.   }

\section{Conclusions}
\label{sec:concl} 
We have presented a scale-based analysis of the bitcoin cryptocurrency
and associated measures of multiscale correlations. 
\rc{
The strong growth spurs in the cryptocurrency can be understood in terms of such correlations and 
``superchaotic'' behavior.  We have shown
that the fractional or anomalous-diffusion behavior that we observe can be attributed
to an inherent temporal correlation structure rather than a non-Gaussian marginal
character.   
 The analysis is carried out using a Haar wavelet based approach which makes it possible to track local changes in the correlation properties of the 
 price.  We find that the changes in the
 price correlation structure can be used to estimate  characteristic epochs
 of relative structural stationarity in the price evolution.  
 We use  the scale spectrum and its parameterization in terms 
 of volatility and Hurst exponent, which
 we view as a market herding index, as a tool to identify four main epochs for the price evolution.
There are two epochs with large Hurst exponents of
approximately $.6$, that are characterized by relatively large price
moves, and in between there is an epoch
 with  Hurst exponent approximately $.5$, that is characterized 
 by a relatively stable price level.   Finally, there is an epoch of relative strong anti-persisitence
 or negatively correlated returns  which corresponds to the recent epoch of price decay.
 The  second epoch of relative efficiency starts after the hacking of the exchange platform
 Mt. Gox and lasts about  $3.5$ years.   }
A main result of our analysis is that over the entire period of bitcoin existence (about seven years)
 the scale spectrum conforms with that of a homogeneous power law 
 with Hurst exponent about $.6$ which
 is larger than  typical values obtained when considering the equity market \cite{eom08}
 or classic currency markets \cite{yao00}.     
 
\rc{We have also introduced a measure of multifractality which is indeed
 relatively strong in the epoch associated with the Mt. Gox hack. 
 We have furthermore introduced a measure of scale-based 
 correlation and have used this to analyze the correlations 
between Bitcoin and two other cryptocurrencies. 
 We see in particular that the correlations of scale-based returns
between bitcoin and  Ripple indicate  a coherence that increases  with 
scale in inefficient markets.}
 
 We remark finally that it is also of interest to look at intraday bitcoin prices.
 Here we have focused on the spectral characteristics of the daily prices, which are 
 important when the time horizon of interest is on the scale of multiple days.
One  can expect to have additional high-frequency 
 intraday spectral features which may be somewhat
 different from those seen for usual currencies and equity markets.
 We do not consider  intraday effects  here,  but we remark that the
 data analytic tools set forth in this paper could be used also for considering
 such intraday spectral features.

 \section*{Acknowledgements}
This work was supported by   in part by 
ul  Lusenn, 
 Centre Cournot, Fondation Cournot,  
 Universit\'e Paris Saclay (chaire D'Alembert).

\bibliography{1901bitcoin.bib}

\appendix

\section{ \rc{Modeling Multi-fractional Brownian Motion}}
\label{app:mBm}
 
The classic model for a random process with power-law behavior 
is fractional Brownian motion  \cite{b18}, whose increments are stationary 
and whose power-law parameters, the Hurst exponent and the volatility, are constant.
Here we present a class of random processes with local power-law behavior, 
whose power-law parameters vary in time. 
This corresponds to a generalization of fractional Brownian motion
to multi-fractional Brownian motion.  
Below we give a precise definition of a multi-fractional Brownian motion and
relate it to our model for the observations. 
Multi-fractional Brownian was introduced in \cite{benassi97,vehel95} and more details can be found in \cite{cohenistas} for instance.
Let $H : \mathbb{R} \to (0, 1)$ and $\sigma : \mathbb{R} \to (0,\infty)$ be two measurable functions.
A real-valued process $B_{H,\sigma}(t)$ 
is called a multi-fractional Brownian motion with Hurst exponent $H$ and volatility $\sigma$
 if it admits the harmonizable representation
\begin{equation}
\label{def:mbm}
 B_{H,\sigma}(t)
 =
\frac{\sigma_t}{\sqrt{C(H_t)}}
{\rm Re}\Big\{  \int_{\mathbb{R}}
 \frac{e^{- i \xi t}-1}{|\xi|^{1/2+H_t}}  d\tilde{W}(\xi) \Big\} ,
\end{equation}
where the complex random measure $d\tilde{W}$ is of the form $d\tilde{W} = dW_1+idW_2$ with 
$dW_1,dW_2$ two independent real-valued Brownian  
measures, 
and $C(h)$ is the normalization function:
\begin{equation}
C(h) =  
\int_\mathbb{R} \frac{4 \sin^2(\xi/2)}{|\xi|^{1+2h}} d\xi 
= 
\frac{\pi}{h \Gamma(2h) \sin(\pi h)} .
\end{equation}

Let $h\in (0,1)$ and $s \in (0,\infty)$. If $H_t \equiv h$ and $\sigma_t\equiv s$, 
then $B^{(h,s)}(t)\equiv B_{H,\sigma}(t)$ is a fractional Brownian motion with Hurst exponent $h$
and volatility $s$, i.e. 
a zero-mean Gaussian process with covariance
\begin{equation}\label{eq:corr}
\mathbb{E} \big[ B^{(h,s)}(t) B^{(h,s)}(t') \big]
=
\frac{s^2}{2}  
\big( |t|^{2h}+|t'|^{2h} -|t-t'|^{2h}\big).
\end{equation}

Let $\beta \in (0,1)$. Let $H : \mathbb{R} \to (0, 1)$ and $\sigma : \mathbb{R} \to (0,\infty)$ be two $\beta$-H\"older functions,
such that $\sup_t H_t < \beta$.
The multifractional Brownian motion (\ref{def:mbm}) is a zero-mean continuous Gaussian process
that satisfies the Locally Asymptotically Self-Similar property  \cite{benassi97}:
At any time $\tau \in \mathbb{R}$,  we have
\begin{equation}
\lim_{\epsilon \to 0^+} {\cal L} \Big(  \big( \frac{B_{H,\sigma}(\tau+\epsilon t) - B_{H,\sigma}(\tau)}{\epsilon^{H_\tau}}\big)_{t\in\mathbb{R}} \Big)
={\cal L} \Big( 
\big( B^{(H_\tau,\sigma_\tau)}(t) \big)_{t\in \mathbb{R}} \Big) ,
\end{equation}
where ${\cal L}$ means ``the distribution of",
which means that there is a fractional Brownian motion with Hurst exponent $H_\tau$ and volatility $\sigma_\tau$
tangent to the multi-fractional Brownian motion $B_{H,\sigma}$. 
This implies that its pointwise H\"older regularity is determined by its Hurst exponent.

Here, we shall assume that the parameter functions $H$ and $\sigma$ are smooth 
with a characteristic time of variation $T_0$ so that we can write
\ban
H_t  = H_0\left(\frac{t}{T_0}\right)  ,  \quad  
\sigma_t  = \sigma_0\left(\frac{t}{T_0}\right)   , 
\ean
where $H_0, \sigma_0$ have variations on the order one scale with respect
to their dimensionless argument. 
Recall from Section \ref{sec:data} that we assume that we consider
the data in a moving window with the $k$th window, $k=1,\ldots,N-M+1$,
having center time $\tau_k$ and width $M\Delta t$.  
In the context of the data in (\ref{eq:data})  we then assume
$M\Delta t \ll T_0 $.
 In distribution  we then model  the daily recorded  log-price recordings in  (\ref{eq:data})  
  by
\ba\label{eq:model} 
 a_0^{(k)}(i)   =  \log\left( P(t_{i+k})\right)  
=     \int_{t_{i+k}-\Delta t/2}^{t_{i+k}+\Delta t/2}   B^{(h,s)}(t)  \, dt ,   \quad i=0,\ldots, M-1,
\ea
with $h=H_{\tau_k}, s=\sigma_{\tau_k}$.
That is,  the data  in   (\ref{eq:data})  are modeled 
 as the level zero approximation coefficients of the  process   $B^{(h,s)}(t)$ relative to the Haar wavelet basis.

\section{Power Law estimation with the Scale Spectrum}\label{app:S}

We explain the details of how   the local power-law parameters are estimated from the log-price data in  (\ref{eq:data})  which are modeled as in  (\ref{eq:model}). 
 
The  input  parameters are the integers $j_{\rm i}<j_{\rm e}$ that determine the scale range under consideration,
the inertial range, and  the window size $M$  which is the size of the moving time window
in   which the local spectra are computed. We must have 
\ba\label{eq:inertial}
1 \leq j_{\rm i} < j_{\rm e} \leq \lfloor M/2 \rfloor .
\ea
  We proceed as follows:

\begin{enumerate}
 \item
Compute the scale spectrum
${\itbf S}^{(k)}=(S_j^{(k)})_{j=j_{\rm i}}^{j_{\rm e}}$ as the local mean square of the wavelet coefficients:
\begin{eqnarray}\label{eq:scs}
S_j^{(k)}=\frac{1}{N_j} \sum_{i=0}^{N_j-1} \big(d_j^{(k)}(i)  \big)^2   ,
\end{eqnarray}
where
\begin{eqnarray}
   N_j         &=&    M - 2j + 1,   \\
   d_j^{(k)}(i)       &=&  \frac{1}{\sqrt{2j}}   \sum_{l=0}^{j-1} 
   a_0^{(k)}(l+i) - a_0^{(k)}(l+i + j)  .   \label{ddef} 
   \label{def:dji}
\end{eqnarray}

\item
Define the $(j_{\rm e}-j_{\rm i}+1)$-dimensional vector ${\itbf Y}^{(k)}$ and  the $(j_{\rm e}-j_{\rm i}+1)\times 2$-dimensional
matrix ${\bf X}$ as 
\begin{eqnarray}
\label{def:Y}
{\itbf Y}^{(k)}  &=&  \big( \log_2(S^{(k)}_{j_{\rm i}}),\cdots,\log_2(S^{(k)}_{j_{\rm e}}) \big)^T, \\
 \label{def:designmatrix}
   {\bf X}   &=&
   \left[
  \begin{array}{cc}
                           1 & \log_2(2 j_{\rm i}) \\
                           1 &\log_2(2(j_{\rm i} + 1) ) \\
                           \vdots & \vdots \\
                           1 & \log_2(  2 j_{\rm e} )
                \end{array}
               \right]  ,
\end{eqnarray}
 and for a  $(j_{\rm e}-j_{\rm i}+1) \times ( j_{\rm e}-j_{\rm i}+1)$-dimensional 
 weighting  matrix~$ {\bf R}$ compute the regression parameters 
 $\hat{\itbf b }^{(k)} = ( \hat{c}^{(k)}, \hat{p}^{(k)})^T$  defined by
\begin{equation}
\label{eq:regb}
\hat{{\itbf b}}^{(k)} = ({\bf X}^T {\bf R}^{-1} {\bf X})^{-1} {\bf X}^T {\bf R}^{-1} {\itbf Y}^{(k)} .
\end{equation}
 In (\ref{eq:regb}) the matrix ${\bf X}$ is the design matrix, ${\bf R}$ is the least squares weighting matrix, and ${\itbf Y}^{(k)}$ is the data vector for the generalized least squares problem that allows
to identify the local power-law parameters. 
\item
Compute the local Hurst exponent and volatility estimates as
\begin{eqnarray}
\label{eq:defhatH}
\widehat{H}(\tau_k;{\bf R})          &=&       \frac{\hat{p}^{(k)}-1}{2}  , \\
\widehat{\sigma}(\tau_k;{\bf R})  &=&       \frac{2^{\hat{c}^{(k)}/2}}{\sqrt{h(\widehat{H}(\tau_k;{\bf R}))}}  .
\label{eq:defhatsigma}
\end{eqnarray}
In (\ref{eq:defhatsigma}) the  scale-spectral scaling function $h$ is defined by
(\ref{hfunc}).

\item
For a diagonal weighting matrix of form (with $q>0$)
\ban
                \label{def:hatR}
{R}^{q}_{j_1j_2} &=& 
 {j_1^q} {\bf 1}_{j_1}(j_2) , \quad j_1,j_2\in\{ j_{\rm i},\ldots,j_{\rm e} \}, 
\ean
compute 
\ban
   \overline{\bf R}^{(k)} = \dessous{\rm argmax}{{\bf R} \in \{ {\bf R}^1,{\bf R}^3 \} } 
   \left( \widehat{H}(\tau_k;{\bf R}) \right) ,
\ean
and obtain the robust parameter estimates by
\ba\label{eq:estsS}
\widehat{H}(\tau_k)   &=& \widehat{H}(\tau_k;\overline{\bf R}^{(k)})   ,\\
\widehat{\sigma}(\tau_k)   &=& \widehat{\sigma}(\tau_k;\overline{\bf R}^{(k)}) .
\ea

\end{enumerate}

We remark that in (\ref{eq:defhatH}) we may threshold the estimation by \\
$\min( \max( \widehat{H}(\tau_k;{\bf R}) ,0.05) , 0.95)$
in order to avoid any singular behavior.   
Note  also that in order to obtain a robust estimator we have used a combination
of two diagonal weighting matrices,  see \cite{GS18} for a discussion
 of the robustness and precision of this procedure.

\section{A Local Measure of Multi-fractality}
\label{app:mom}

In \cite{bit_cite} the authors use a multi-fractal detrended fluctuation analysis (MF-DFA \cite{mdfa})
to detect multi-fractality. For a given duration $s$, this amounts
to divide the time series into equal subsections of duration $s$ and 
 to subtract from the time series a fitted polynomial of order $m$ for each subsection.
Then, the variance of the detrended data is computed for each 
subsection. If the total
data length is $N$ then the variances of the detrended data in each subsection are denoted by
\ban
F^2(s,v), \quad v=1,\ldots, N_s=\lfloor N/s  \rfloor. 
 \ean
For a general moment $q>0$  
an  analogue of the scale  spectrum is formed as
\ban
    F_q(s)  = \left\{ \frac{1}{N_s} \sum_{v=1}^{N_s} F^2(s,v)^{q/2}  \right\}^{1/q}   .
\ean
A linear regression according to the model 
\ban
   \log\left( F_q(s)  \right)    =   \log(A)   +  H(q) \log(s)  
\ean 
is carried out to get a generalized Hurst exponent $H(q)$.  
In the mono-fractal case the 
generalized Hurst exponent $H(q)$ should be $q$-independent
(since $F^2(s,v)$ should be independent of $v$) and it should be equal to the actual Hurst exponent.
In \cite{bit_cite} the authors  carry out such an analysis for each
of the low and high price epochs (before and after February 2013) 
for both the bitcoin prices and returns.
They find a stronger degree of multi-fractality in the first epoch,  as
reflected  by a stronger dispersion in the generalized Hurst exponent.
This is the case for both the returns and prices, moreover, the authors
establish (by a randomly shuffling method) 
that fat-tails are  the main source of multi-fractality in both low- and high-price 
epochs (i.e. before and after February 2013).

Here, we carry out a related analysis with a modified 
generalized Hurst exponent. First, we want to track the actual
multi-fractal character, the changes 
in the Hurst exponent,   that is why we use a moving window 
rather than a ``global''  analysis. Second, we want to look at the scaling
 structure for all scales and therefore do not subtract a fitted trend polynomial.
 
In order to detect residual multi-fractality  within each window we 
form the generalized scale spectrum by  
 \begin{eqnarray}
\label{scale-sp:p}
S_j{(q)}= \Big\{ \frac{1}{N_j}\sum_{i=0}^{N_j-1} |d_j(i)|^q \Big\}^{2/q} ,
\end{eqnarray}
where we suppress the dependence on the window (with center point $\tau_k$). 
When the underlying process is
fractional Brownian motion with Hurst exponent $H$ we have 
(provided $N_j \gg 1$ so that $S_j{(q)} \approx \EE[|d_j(1)|^q]^{2/q}$): 
 \begin{equation}
 \label{eq:expressHgenq}
  \log_2\big( S_j{(q)}  \big)  \approx   
 \frac{2}{q}   \log_2 \Big(  \frac{2^{q/2}}{\sqrt{\pi}} \Gamma\big(\frac{q+1}{2}\big)\Big)
+   \log_2(\sigma^2 h(H)) +j (2H+1)   .
\end{equation}
For an arbitrary process we get a generalized
Hurst exponent, $H(q)$,
 via linear regression with respect to this model:
 $$
  \log_2\big( S_j{(q)}  \big)   = \log_2 (A)  + j (2 H(q)+1).
 $$
By (\ref{eq:expressHgenq}) this generalized exponent  is $q$-independent when the underlying process is 
 mono-fractal Brownian motion. 

In Figure  \ref{fig:moment} we show $H(q;\tau_k)$ for the bitcoin data
as function of window center point $\tau_k$.
We can see some degree of within window multi-fractality 
in the first epoch of high Hurst exponent in early 2013, moreover,
a rather high degree of within window multi-fractality just after 
the hacking of Mt. Gox.  However, in general the degree of 
within window multi-fractality is relatively low.  We remark that,  
for a finite window width there is a $q$-dependent
bias in the generalized Hurst exponent estimate (as can be shown by a direct calculation for 
fractional Brownian motion), thus to examine the stability of the
multifractal variations  and eliminate this bias we center the curves
with respect to the mean of the curve with $q=2$.

\section{Tail and Marginal Effects}
\label{app:marg}
 
\begin{figure}[htbp]
   \centering
      \includegraphics[width=6.0cm]{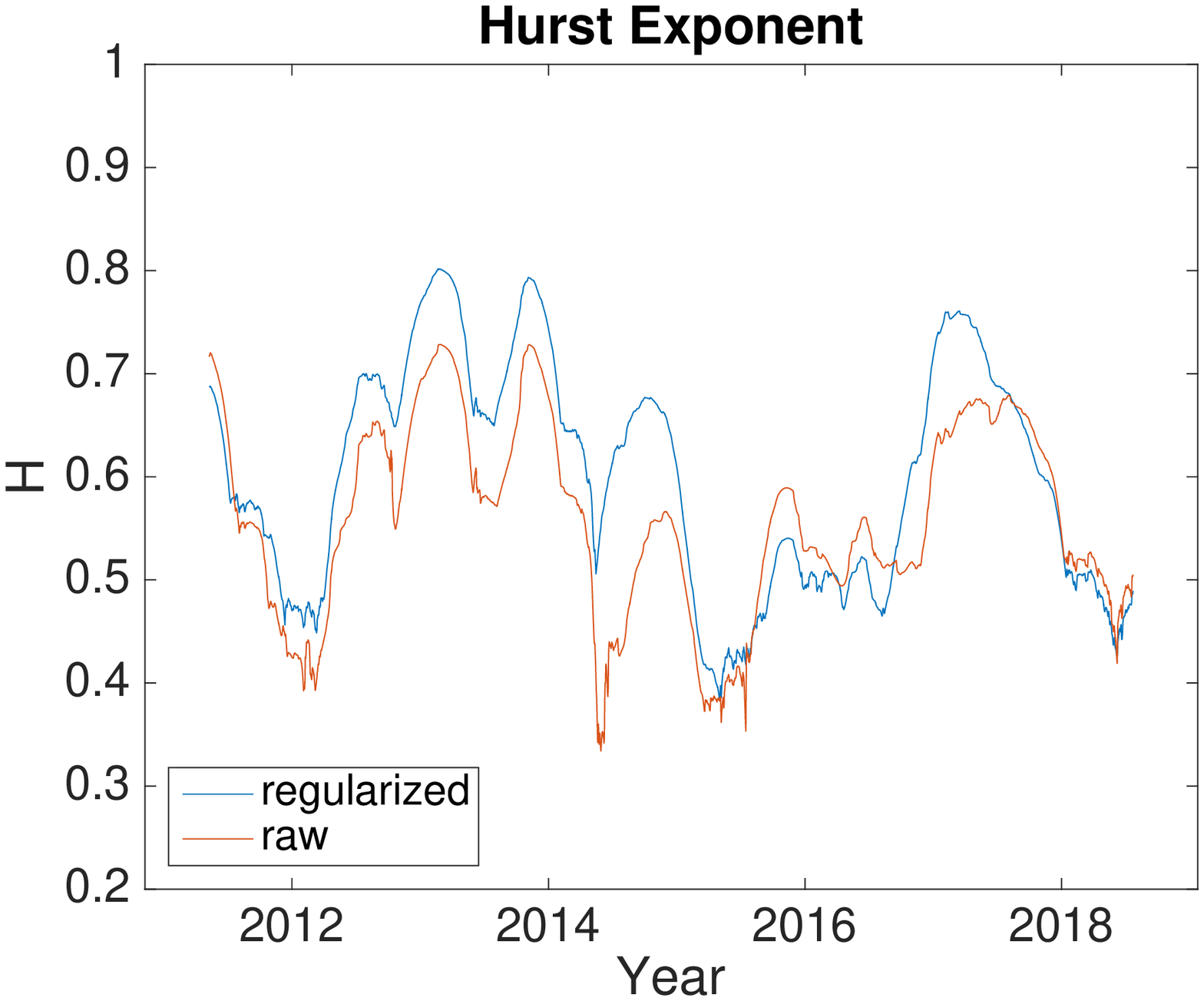}  
     \includegraphics[width=6.0cm]{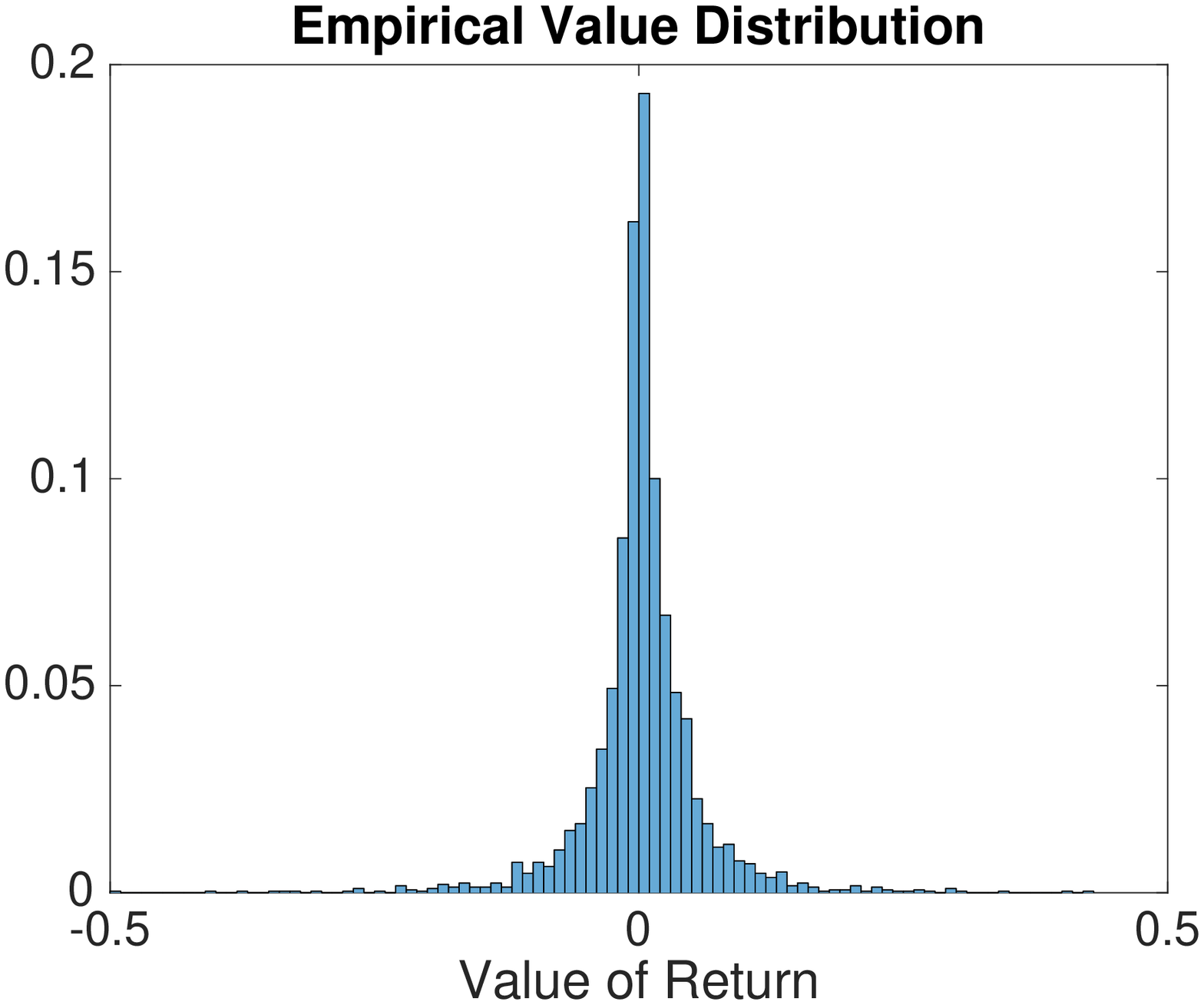}  
   \caption{Log-price  differences (left) over the entire period of bitcoin existence and histogram (right).
   }
   \label{fig:m1}
\end{figure}

\begin{figure}[htbp]
   \centering
      \includegraphics[width=6.0cm]{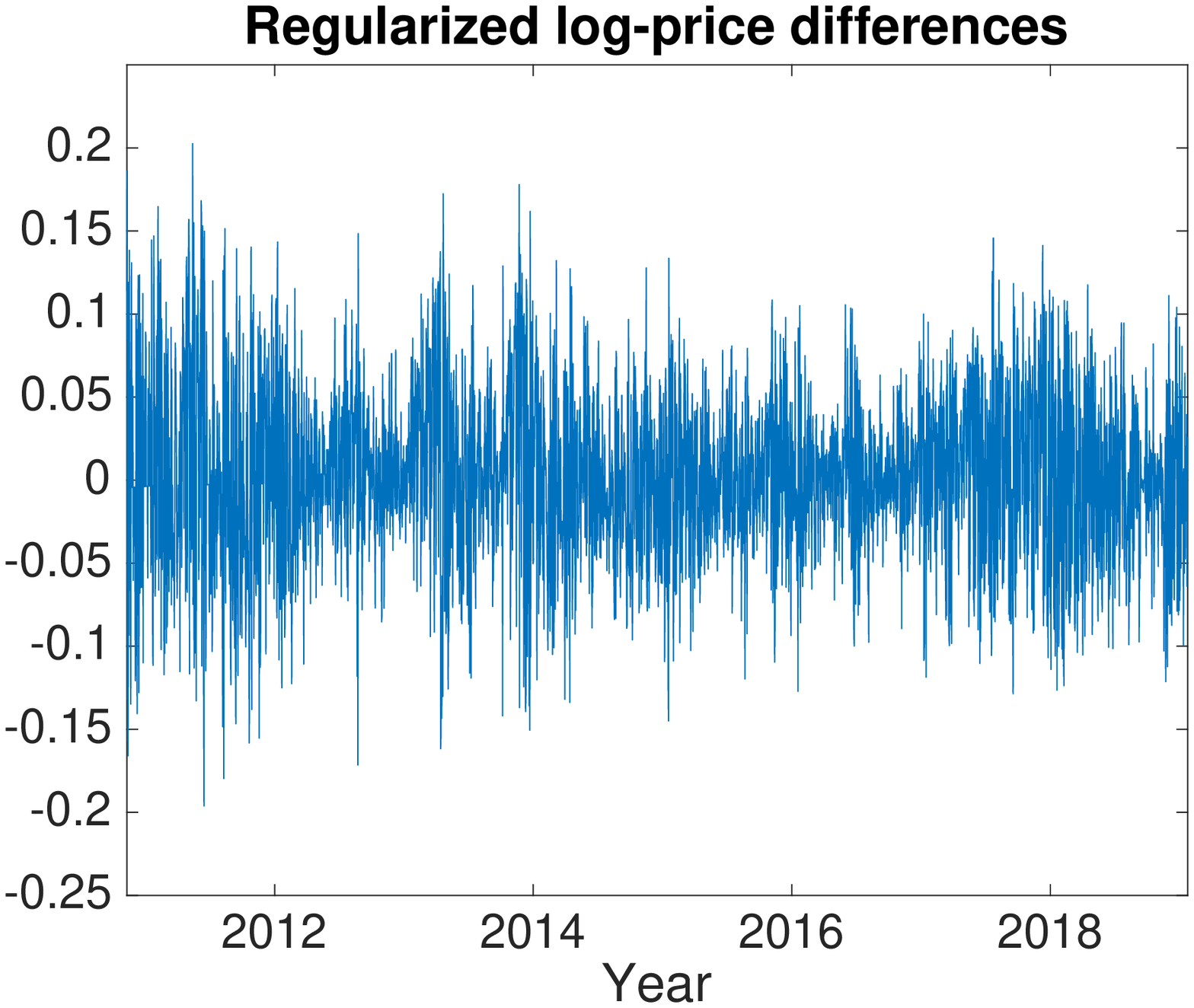}  
      \includegraphics[width=6.0cm]{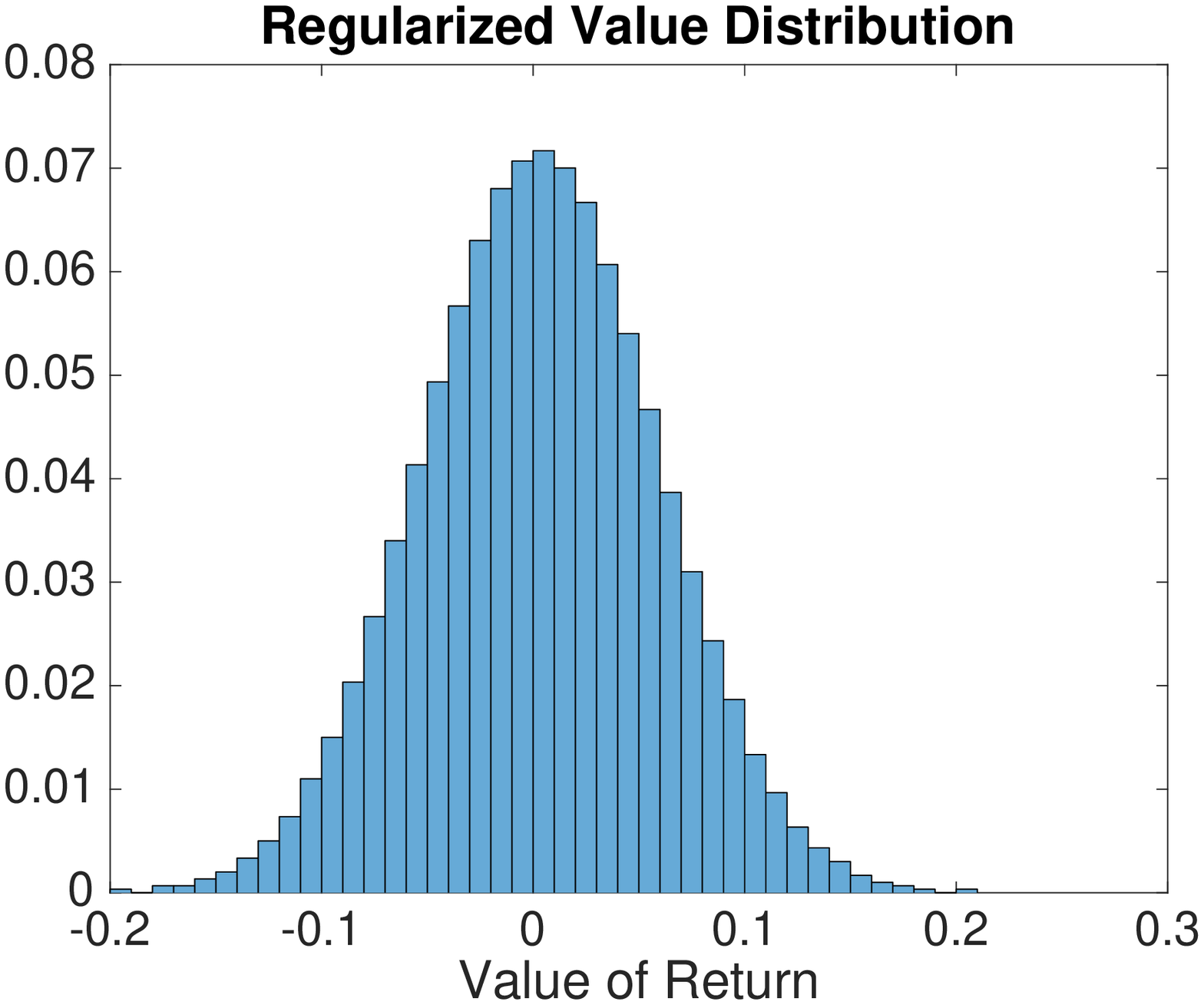}  
   \caption{ Regularized log-price  differences (left) over the  entire period of bitcoin existence  and histogram (right).
  }
   \label{fig:m24}
\end{figure}

\begin{figure}[htbp]
   \centering
      \includegraphics[width=6.0cm]{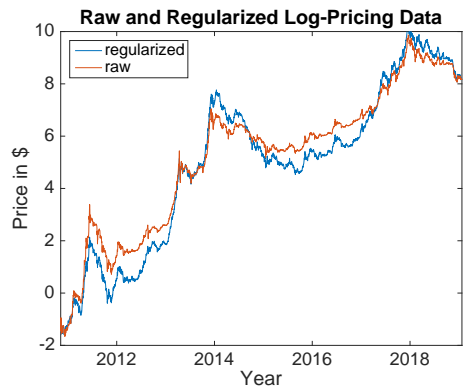}  
   \caption{ Raw (red) and regularized (blue) log prices.
   }
   \label{fig:m25}
\end{figure}

\begin{figure}[htbp]
   \centering
      \includegraphics[width=6.0cm]{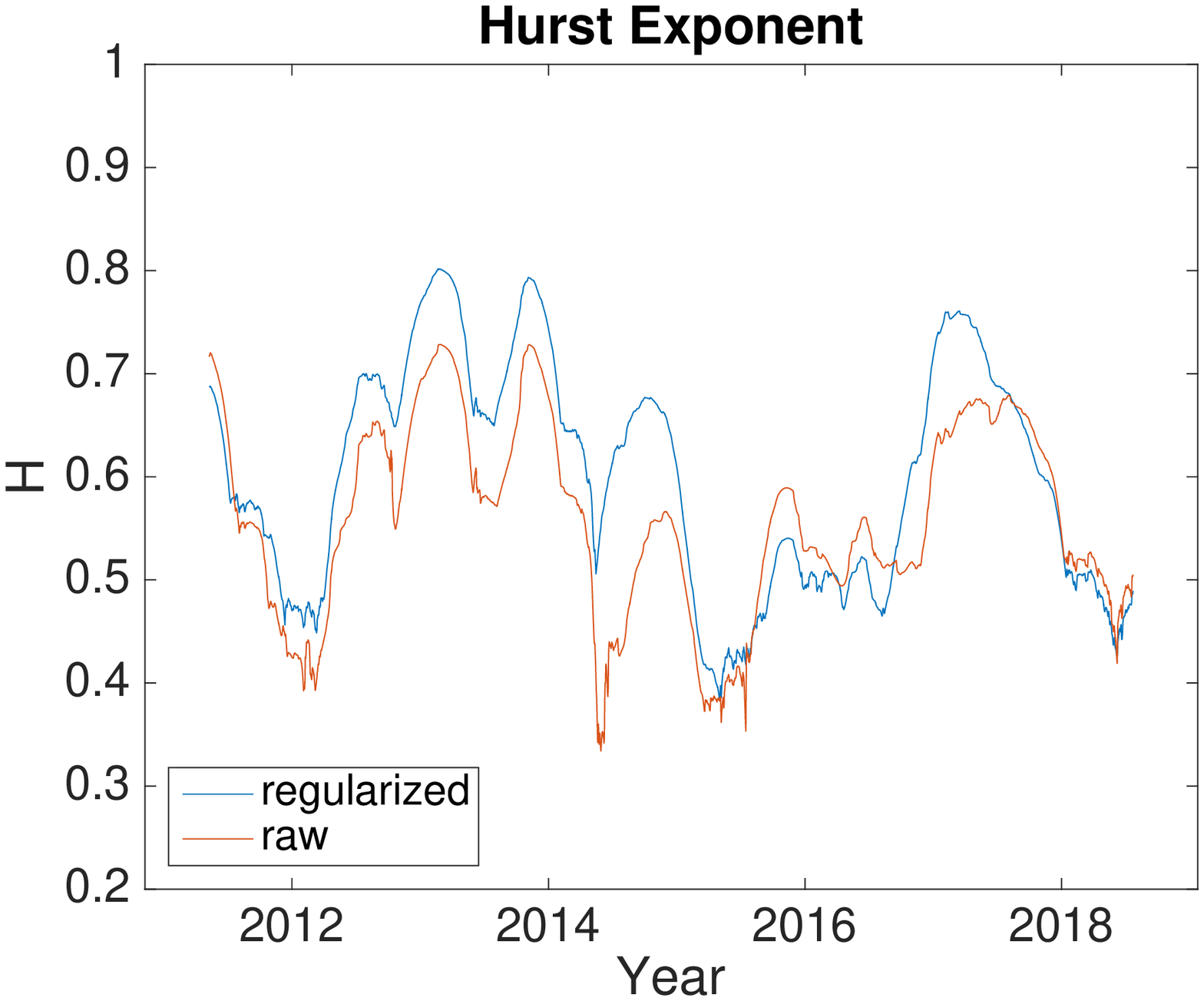}  
      \includegraphics[width=6.0cm]{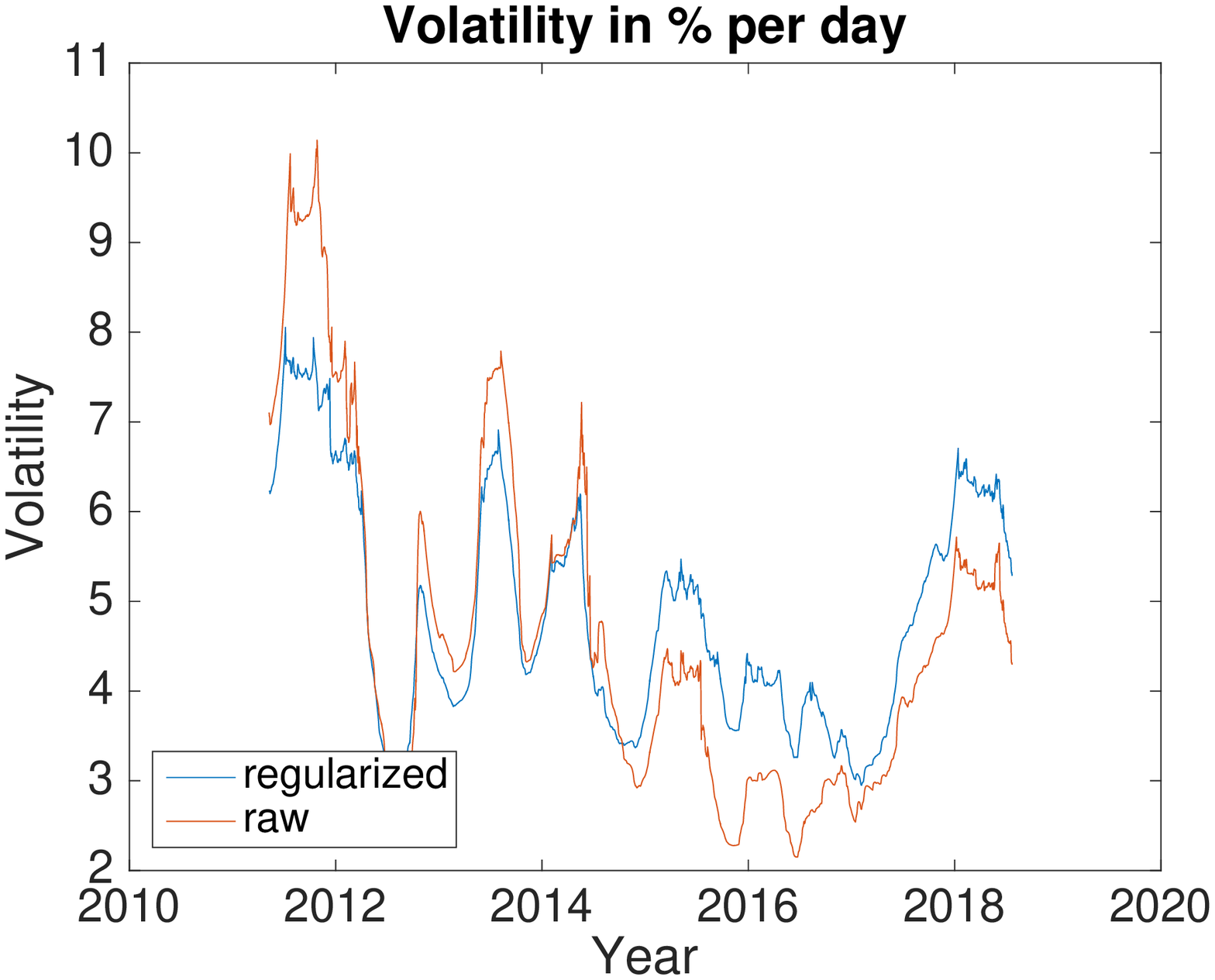}  
   \caption{ Hurst exponent (left) and daily volatility (right) of the raw returns (red) and the regularized returns (blue).
   }
   \label{fig:m27}
\end{figure}

 In this appendix we show that the fluctuations in the Hurst exponent,
 as a measure of multi-fractality, are not due to tail or marginal non-Gaussianity
 effects.
The raw log-price  differences are shown in Figure \ref{fig:m1}.
The histogram exhibits a heavy-tail distribution.
 We regularize by a Gaussian marginal 
 transformation of the log-price differences.
 Note that the log prices are differentiated, then the transformation is carried out (see Figure \ref{fig:m24}),
 and the regularized log-prices created by integration (see Figure \ref{fig:m25}) and subsequently 
 used in the multi-scale analysis as before.
 We can see in Figure \ref{fig:m27} that the power-law parameters are essentially 
 the same ones for the raw data and for the regularized (Gaussianized) data.     
 The only minor change is that the procedure applied to the regularized data seems to slightly overestimate large $H$ values compared to the case where the raw data are processed directly.  
Other regularization methods (such as truncation at plus or minus two standard deviations)
give the same result.

\section{On Chaotic Behavior}
\label{app:chaos}
Following \cite{bit_cite}  we can in the simplest case of a linear
homogeneous system of embedding dimension one consider a 
chaotic system by
\ban
   X_{n+1}  = a X_{n} +    \sigma \epsilon_n, \quad X_0=x_0 ,
\ean
with  $ \epsilon_n$ a sequence of independent and identically distributed standard Gaussian noise terms.
The associated  Lyapunov exponent is
$\lambda = \log(a) $.
If  $\lambda > 0$ then we have a chaotic system and an exponential growth:
\ban
    {\rm St.Dev}\left[ X_n \right]  =  
    \sigma \sqrt{ \frac{e^{2\lambda n}-1}{e^{2\lambda} - 1 } }   . 
\ean

Above we found from the global spectrum, see Figure
 \ref{fig:global},  that 
the effective returns in the Gaussian case could be modeled
as increments of fractional Brownian motion with Hurst
exponent $H=.6$. In the continuous limit we may then
model the price  process by
\ban
   P(n\Delta t) = p_0  \exp\big( \sigma B^H(n\Delta t) \big) ,
\ean 
and we have
\ban
   {\rm St.Dev} \left[ P(n\Delta t) \right] = p_0   \exp\big(  \sigma^2 |n \Delta t|^{2H} \big)  
   \Big(1 - \exp\big( - \sigma^2 |n \Delta t|^{2H} \big)  \Big)^{1/2} .
\ean
Thus, the price process, with $H=.6$, grows at super-exponential rate
or  is ``super-chaotic''. In the anti-persistent case with $H<1/2$
 the price process  grows at sub-exponential rate. 
Note also that  the return  process itself  is fractional Gaussian
noise, and thus exhibits long-range correlations, however,
is stationary and does not exhibit a chaotic behavior.

  \end{document}